\documentclass[12pt]{article}

\pdfoutput=1

\usepackage{subfiles}
\usepackage{graphicx,color}
\usepackage{caption}

\begin{document}

\title{Self-assembly of colloid-cholesteric composites provides
a possible route to switchable optical materials}

\author{K. Stratford$^{*}$, O. Henrich$^{*}$, J.~S. Lintuvuori$^{*}$,\\
M. E. Cates, D. Marenduzzo$^\dagger$ \\ 
SUPA, School of Physics and Astronomy, The King's Buildings,\\
The University of Edinburgh, Edinburgh, EH9 3JZ, UK\\
\small{$^{*}$ equally contributed to this work}\\
\small{$^\dagger$ Corresponding author: dmarendu@ph.ed.ac.uk}
}
\date{}

\maketitle

\noindent

\noindent
{\bf
Colloidal particles dispersed in liquid crystals can form new materials
with tunable elastic and electro-optic properties. In a periodic `blue
phase' host, particles should template into colloidal crystals with
potential uses in photonics, metamaterials, and transformational optics.
Here we show by computer simulation that colloid/cholesteric mixtures can
give rise to regular crystals, glasses, percolating gels, isolated clusters,
twisted rings and undulating colloidal ropes. This structure can be tuned
via particle concentration, and by varying the surface interactions of the
cholesteric host with both
the particles and confining walls. Many of these new materials are metastable:
two or more structures can arise under identical thermodynamic conditions.
The observed structure depends not only on the formulation protocol, but also
on the history of an applied electric field. This new class of soft materials
should thus be relevant to design of switchable, multistable devices for
optical technologies such as smart glass and e-paper.
}

\bigskip
\noindent

\noindent
When spherical colloidal particles are mixed into a nematic liquid crystal, 
they disrupt the orientational order in the fluid and create defects (disclination lines) in the nematic close to their surfaces. To minimise the free energy cost, it is generally advantageous for defects to be shared:
therefore the colloidal particles have a generic tendency to aggregate. Such aggregates may further 
self-assemble into lines~\cite{wiresmiha}, 2D crystals~\cite{zumer}, 
planar structures~\cite{tanaka}, or 3D amorphous glasses~\cite{tiffany}.
Besides being of fundamental interest to materials science, these
structures may have tunable elastic and optical properties~\cite{stark}. Hence 
they offer exciting prospects for applications as biosensors~\cite{abbott}, or
as new devices~\cite{colloiddevice,tanakanatmat}.

But one may also choose to disperse colloidal particles in a cholesteric, 
rather than nematic, liquid crystal. The molecules making up this material are
chiral, and this causes their average orientation (described by a director field) to rotate in space in
a helical fashion, rather than remain uniform as in nematics. 
This typically creates a 1-dimensional periodic structure, called the cholesteric phase, whose wavelength 
is the pitch, $p$. The periodicity can however become fully three dimensional in the so-called blue phases (BPs)~\cite{mermin}. BPs arise because the director field can twist around more than one direction at once (see 
the cartoon in Supplementary Fig.~1). Such double twist 
regions (generally cylinders) are energetically favoured at high enough molecular chirality, but
lead to geometric frustration because it is impossible to tile the whole 3D
space with double twist cylinders without also introducing ``disclination lines'' 
(These are singular topological defects on which the director field is 
undefined.) In blue phases these disclination lines themselves either form a 3D periodic regular lattice (in so called BPI and BPII), or remain disordered (BPIII)~\cite{bp3}. 
BPs are stabilised by the tendency of chiral molecules to twist, but destabilised by the cost of forming topological defects (which locally diminish the molecular alignment). In conventional cholesteric materials~\cite{mermin} this balance is only achievable in a narrow temperature range of a few degrees, around the onset of liquid crystalline ordering.
However, recent advances in formulation have widened this stability
range enormously~\cite{kikuchi,bplasers,coleswidetrange,bpdevice}. Most
recently, templated BPs with a stability range of -125
to~125$^\circ$C have been reported \cite{coles}, paving the way to
applications of BPs in operational display devices.

Besides being remarkable materials in themselves, BPs offer significant promise as hosts for dispersing colloidal particles. Because BPs contain a disclination network even in the absence of
particles, they can potentially template colloidal self-assembly. 
This was proved recently by simulations of BP-dispersed nanoparticles which 
exclude the surrounding liquid crystal but otherwise have negligible surface 
interaction with it (to give a so-called `weak anchoring' regime)~\cite{extrareference1,miha}. 
Such particles are attracted to the disclination network in BPs: by covering up the defect cores, the high elastic energy cost of those regions is avoided. The pre-existing order of the defect network then templates particles into a regular colloidal crystal of the same periodicity. 
Because this structure has a wavelength in the visible range, 
the resulting material should inherit (and probably enhance) the incomplete photonic bandgap of the parent blue phase, as exploited for instance in laser devices~\cite{bplasers}; for related possibilities see \cite{lavrentovich}.

A second compelling reason to study colloidal-BP composites is that, even without particles, BPs can show several competing metastable free energy minima,
corresponding to different topologies of the defect network~\cite{adriano,fukuda}, with a strong dependence on applied electric or magnetic fields~\cite{henrichfield}.
Adding particles is likely to further enrich the free energy landscape, creating composites that might be promising candidates for bistable or multistable devices, in which energy is needed only to switch optical properties and not to maintain them. (This is the e-paper paradigm~\cite{epaper}, and can lead to huge energy savings.)
To explore such metastability, and switching strategies between states, we require multi-unit-cell, time-dependent simulations. These go beyond previous studies of colloidal-BP composites~\cite{miha} but follow comparable simulations of pure BPs~\cite{bp3,henrichfield,domaingrowth}, and isolated and dimeric colloidal particles in cholesterics~\cite{juho1,juho2}.

The simulations presented here are representative of a liquid crystal
blue phase with unit cell size in the range of 100--500~nm \cite{mermin},
and particles with a diameter of around 50~nm. The simulations
address both the bulk structure, and the structure which forms when
the colloidal dispersion is confined in a narrow sandwich of width
comparable to the unit cell size~\cite{extrareference2}. It is seen
that a wide variety of structure is possible, and may be influenced
by parameters including the solid volume fraction, and the details
of the anchoring at the colloid surface (and the at the surface of the
confining walls). Further, evidence is found for metastable switching of
this structure in an applied electric field. Details of the
computational approach are set out in the Methods section, while
simulation parameters are reported in the Supplementary Methods.

\bigskip
\noindent
\textbf{\large Results}

\noindent
\textbf{Structure in bulk BPI.}
We show in Fig.~1~(\textbf{a}--\textbf{d})
the final state structures of four simulations in which we first equilibrated a stable BPI disclination network in bulk (with periodic boundary conditions) and then dispersed colloidal particles at random within it. Among dimensionless control parameters are the particle solid volume fraction $\phi$  ($\phi = 1\%$ or $5\%$); the ratio $r = R/\lambda$ of particle radius to the BP lattice parameter $\lambda$ (here $r\simeq 0.14$) and the ratio $w = WR/K$ where $W$ is the anchoring strength of the cholesteric at the particle surface, and $K$ is the elastic constant of the liquid crystal.
The anchoring strength is defined such that $W > 0$ imparts a preferred
orientation to the director field at the particle surface. This orientation
can be either normal to the surface or in a plane tangential to the surface,
a property related to surface chemistry in experimental systems.
We choose $w = 0.23, 2.3$ which are within typical experimental range~\cite{tiffany}. This parameter can be viewed as the ratio of the anchoring energy, $WR^2$, to the elastic energy scale for distortion~$KR$.

For a single colloidal particle in a nematic fluid, $w$ controls the formation of a local topological defect at the particle surface~\cite{stark}, with weak distortion of the nearby fluid at small $w$. In the colloidal-BP system, $w$ also plays a determining role in the final composite structure. When $w$ is small
(Fig.~1\textbf{a} and Fig.~1\textbf{b}; Supplementary Movies~1 and~2),
the attraction of particles to the disclination lattice causes
 the lattice to become covered by particles, with little disruption of its
 long-range order. Thus a templated colloidal crystal is created, as first reported in~\cite{miha}. Interestingly, the weak but finite anchoring apparently creates an elastic force between the colloidal particles which leads to very slow dynamics
(Supplementary Fig.~4) and the formation of small colloidal lines within disclinations, especially for dilute samples (Fig.~1\textbf{a}). It seems likely that polymer-stabilised  blue phases~\cite{kikuchi} work by a similar principle to this small $w$ case, with a weak segregation of monomers towards the defect regions.

For large $w$ (Fig.~1\textbf{c} and Fig.~1\textbf{d};
Supplementary Movies~3 and~4), the picture changes
dramatically.
Each colloidal particle now disrupts significantly the order of the nearby fluid, with defect formation close to its surface~\cite{stark}. The resulting strong disruption of local liquid crystalline order around each particle leads now to 
a strong interparticle attraction. This restructures the disclination
network completely, destroying the long range order of
the original BPI topology. Most notably, particle aggregation favours
the formation of defect junctions, where four disclinations meet. This structural motif is present in the unit cell of BPII rather than BPI (and also seen in the amorphous BPIII~\cite{bp3}). The particle clusters are disjoint at low volume fractions ($\phi\simeq 1\%$) but they interact elastically via the connecting disclinations. At larger $\phi$, the aggregates join up to form a percolating colloidal cluster which essentially templates the disclinations rather than vice versa. 

As colloidal-nematic composites~\cite{tiffany}, all the structures in
Fig.~1~(\textbf{a}--\textbf{d}) 
should be soft solids, of nonzero elastic 
modulus $G$ at low frequency. This contrasts with pure BPs, where
$G=0$ as the fluid can flow with a finite
viscosity via permeation of molecules through a fixed defect lattice~\cite{permeation1,permeation2}. We expect the
structures in Fig.~1~(\textbf{a}--\textbf{c}) to be only weak gels,
as here the 
defect network percolates but particle contacts do not. This is broadly similar to the cholesteric-nanoparticle gel reported
in~\cite{lubensky} (with $G<1$ Pa). In contrast, the colloidal gel
structure in Fig.~1\textbf{d} should be much more resistant mechanically, 
due to the formation of a thick percolating particle network. The 
structure is now akin to that of the self-quenched glass found in 
colloidal-nematic composites~\cite{tiffany} at much higher volume 
fraction (over 20\% as opposed to below 5\% here), for which $G\sim 10^3$ Pa.
This figure is consistent with an estimate obtained by comparing
the mechanical properties of the network of defects stabilised by
colloidal particles with those of rubber~\cite{ramos} for
the relevant simulation parameters (see Supplementary Table~1).

A more practical protocol for including particles in a BPI phase is to
disperse the particles at random in an isotropic phase, and then
quenching this into the temperature range where BPI is stable.
Here we know that, without particles, the kinetics can favour
disordered (BPIII-like) intermediates which may be long lived
or even metastable~\cite{domaingrowth}. Fig.~1~(\textbf{e}--\textbf{h})
shows outcomes of such a quench protocol with the same parameter sets as
in Fig.~1~(\textbf{a}--\textbf{d}). For small $w$, the particles again
decorate the disclination lines, but are this time templated onto an
amorphous network rather than an ordered one.
Increasing the volume fraction (Fig.~1\textbf{f}) leads to a more even
coverage with a better defined particle spacing. 
For large $w$, on the other hand, strong interparticle forces again dictate
the self-assembly. At low concentrations, discrete particle clusters
decorate an amorphous defect lattice (Fig.~1\textbf{g}) whereas at
higher $\phi$ a percolating network of colloidal strands
(the gel in Fig.~1\textbf{h}) is again seen. 
The structures in Fig.~1~(\textbf{a}--\textbf{h}) have been followed for
$\sim$20 ms 
(see Supplementary Fig.~4); they correspond to either very slowly evolving
states or  deep metastable minima (elastic interactions are of
order $KR$ and dwarf thermal motion, see
Supplementary Fig.~4 and Ref.~\cite{stark,lavrentovich}).

It may be possible to anneal the structures obtained via the quench route,
although we have not attempted this in simulation. The results presented
suggest that even when starting with colloidal particles randomly dispersed
in an ordered blue phase, the particles are able to distort the structure
(Fig.~1~\textbf{a},\textbf{b}). This suggests that it would be difficult
to identify a quench
protocol which would lead to less, rather than more, disorder.

\medskip
\noindent
\textbf{Structures in confined BPI.}
In display devices and other optical applications, it is typical for liquid crystals to be strongly confined (and subject to strong boundary effects), in contrast to the bulk geometries considered above. We therefore next describe how colloidal suspensions self assemble in thin sandwiches (thickness $h = 1.75 \lambda$), whose confining walls can favour either tangential (planar degenerate) or normal (homeotropic) orientation of the director field. For simplicity we here impose one or the other as a strict boundary condition.

Fig.~2~(\textbf{a}--\textbf{d}) shows the case of normal wall anchoring on
varying $\phi$ and $w$. We quench from the isotropic to the ordered phase
as was done in Fig.~1~(\textbf{e}--\textbf{h}), and again find a metastable,
amorphous disclination network
(see Supplementary Methods). The resulting colloidal-cholesteric composites
are somewhat similar to those seen in the bulk: for small $w$ the particles
are templated by the network (Fig.~2\textbf{a},\textbf{b}), whereas for
large $w$ we observe isolated clusters (Fig.~2\textbf{c}) and percolated
colloidal gels (Fig.~2\textbf{d}), for $\phi = 1\%$ and $4\%$ respectively.
There are, however, significant effects of the confining walls.
First, the normal anchoring at the wall recruits particles there,  giving
(for the chosen $h$) a density enhancement of the colloid particles at
both walls  (see side-views in Fig.~2\textbf{a}--\textbf{d}). Second,
the clusters formed at large $w$ are themselves anisotropic with an
oblate habit. These clusters
are similar to those reported experimentally for large colloid particles in a
cholesteric (but non-BP) host~\cite{niek}.
 
Fig.~2~(\textbf{e}--\textbf{h}) shows comparable simulations for planar wall
anchoring.  Without colloidal particles, planar anchoring results in a
regular cholesteric helix,
with axis perpendicular to the boundary walls. 
This twisted texture develops via a set of defect loops,
which shrink and disappear to leave a defect-free sample. However,
dispersing particles with small $w$ arrests the loop shrinking process,
stabilising twisted rings (Fig.~2\textbf{e}) at small $\phi$, and more
complicated  double-stranded structures at larger $\phi$ (Fig.~2\textbf{f}).
For larger $w$, we observe thick colloidal ``ropes" about 5 particles across
(Fig.~2\textbf{g},\textbf{h}). These ropes undulate to follow the chiral
structure of the 
underlying fluid. On increasing the density they develop branch points and
finally again approach a percolating gel morphology. Note that planar wall
anchoring is incompatible with the normal anchoring on the colloidal
particles (for $w>0$), which are thus repelled from both walls
(see the side views in Fig.~2\textbf{e}--\textbf{h}).

The degree of similarity between
the bulk and confined structures is somewhat masked by the local
change in particle density in the confined case. If one allows for
this, Figs.~1 and~2 show the individual structural motifs are actually
rather similar: templated lines for weak anchoring and clusters for
strong anchoring.

\medskip
\noindent
\textbf{Switching in electric field.}
An interesting question, relevant for applications to devices, is whether 
these structures can be modified, or switched, with electric fields.
Fig.~3 addresses this question, starting from amorphous blue phase~III,
this time at a higher chirality. Dispersing colloidal particles in this
network
results in the particle occupying the available nodes, where the order
parameter is low (Fig.~3\textbf{a}; Supplementary Movie~5).
Applying a high enough field 
along the $x$ direction leads to a switching to a new phase, where colloidal 
arrays fit within a network with distorted hexagonal cells (Fig.~3\textbf{b} 
and Supplementary Fig.~7; Supplementary Movie~6). 
Remarkably, upon field removal this honeycomb structure does not find its 
way back to the 
configuration before switching on the field, but gets stuck into a metastable 
structure (Supplementary Fig.~7), with residual anisotropy along the field
switching 
direction. Cycling the field along the $x$ and $y$ directions leads to
distinct metastable structures, which are retained after field removal
(Supplementary Fig.~8 and Fig.~9).
We note that for the fluid parameters used here, the field induced
changes are irreversible with the weakly anchoring colloid particles.
The mesophase without colloid particles undergoes
a reversible ordering transition~\cite{bp3}; the reversibility
is retained in the
presence of particles only if their anchoring is very weak.

\medskip
\noindent
\textbf{\large Discussion}

\noindent
The surprising wealth of colloidal-BP composites obtained in Figs.~1-3 is
the complex consequence of a simple competition between two self-assembly
principles. 
First, blue phases provide a three-dimensional template onto which
colloidal particles are attracted so as to reduce the elastic stresses
arising at the disclination cores~\cite{miha}. Second, colloidal intrusions
in a liquid crystal raise the free energy locally; sharing this cost creates
a generic tendency to aggregate into clusters~\cite{tiffany}. Which of these
factors dominates depends on $w$, a dimensionless measure of the anchoring
strength of the director field at the colloidal surface. For $w\ll 1$, the
templating principle dominates. Colloidal crystals, gels or twisted rings
can then arise. For $w \gg 1$, interparticle attractions defeat the
templating tendency, and we predict disconnected aggregates, percolating
gels, or helical colloidal ropes. In both cases, control is offered by the
process route, colloidal concentration, and the anchoring conditions at
sample walls. We have also shown that electric fields can be used to
switch between metastable states, providing a possible route to future
multistable device applications.

How do these simulations relate to experiment? Several authors have
used nanoparticles in the context of stabilising BPs
\cite{yoshida2009,karatairi}. These
nanoparticles are typically a few nanometres in size, in which case
stabilisation is explained by the mechanism of defect removal.
Experiments suggest this mechanism is also
relevant for larger particles in the range of 40--70~nm 
\cite{dierking}, which is consistent with the current simulations.
However, the mechanism is observed to be less efficient for
particles above 100~nm, where particles may start to disrupt
longer-range liquid crystal order \cite{dierking}. While the field of
self-assembly with nanoparticles is relatively new \cite{draper},
nanoparticles have also been used in both nematic \cite{milette2012}
and chiral liquid crystals \cite{cordoyiannis} as host fluid templates.
The variation of structure seen
in our simulations suggest that some careful characterisation of parameters
such as surface anchoring of particles (which are often stabilised, or even
functionalised with surface ligands), together with those associated
with any confining surfaces or interfaces, is required to understand the
final structure and to be able to reproduce it with a given experimental
protocol. The identification of a model system which could allow
unambiguous comparison between experimental systems and simulation
would be extremely useful. Such a system might adopt particles of
diameter in the range 50~nm or above, which allows a coarse-grained simulation
approach of the type used here to be adopted, which can then capture the
large-scale structure required to compare with experiment. In the same vein,
detailed surface chemistry is difficult to represent in coarse-grained
simulations, so heavily functionalised particles might be avoided.

We have not explored in this work switchability by flow \cite{flowswitch}, but this could again be inherited from the underlying BP dynamics \cite{adriano,permeation2}, in which case the same materials may also be relevant to the emerging technology of optofluidics~\cite{optofluidics}.

\bigskip
\noindent
\textbf{\large Methods}

Here, the simulation method is described. Note that
all quantities are expressed in simulation units. Details
of the simulation parameters, and their mapping to physical units,
may be found in Supplementary Table~1 and the Supplementary Methods.

\medskip
\noindent
\textbf{Thermodynamics.}
The thermodynamics of the cholesteric liquid crystal can be described
by means of a  Landau-de Gennes free energy functional $\cal F$,
whose density is written as $f$:
\begin{eqnarray}
{\cal F}[{\bf Q}]&=&\int d^3{\bf r} f({\bf Q}({\bf r})).
\end{eqnarray}
This free energy density $f = f({\bf Q})$ may be expanded in powers of the
order parameter ${\bf Q}$ and its gradients; ${\bf Q}$ is a traceless 
and symmetric tensor which is denoted from now on in subscript
notation as $Q_{\alpha\beta}$.
The largest eigenvalue $q$ and corresponding eigenvector $n_\alpha$
of $Q_{\alpha\beta}$ describe the local strength and major orientation axis
of molecular order.
The theory based on the tensor $Q_{\alpha\beta}$, rather than a theory based
solely
on the director field $n_\alpha ({\bf r})$, allows treatment of disclinations
(defect lines) in whose cores $n_\alpha$ is undefined.

Explicitly, the free energy density is:
\begin{eqnarray}
f(Q_{\alpha\beta}) &=& {\textstyle \frac{1}{2}} A_0
 \left(1- {\textstyle \frac{1}{3}}\gamma \right)Q^2_{\alpha \beta}
-{\textstyle \frac{1}{3}}A_0\gamma Q_{\alpha \beta} Q_{\beta \gamma}Q_{\gamma \alpha} 
+ {\textstyle \frac{1}{4}} A_0\gamma (Q^2_{\alpha \beta})^2  \nonumber\\
&+& {\textstyle \frac{1}{2}} K(\varepsilon_{\alpha \gamma \delta}
\partial_\gamma Q_{\delta \beta} + 2 q_0 Q_{\alpha \beta})^2
+ {\textstyle \frac{1}{2}} K (\partial_\beta Q_{\alpha \beta})^2.
\label{free}
\end{eqnarray}
Here, repeated indices are summed over, while terms of the form
$Q^2_{\alpha\beta}$ should be expanded to read $Q_{\alpha\beta}Q_{\alpha\beta}$;
$\varepsilon_{\alpha\gamma\delta}$ is the permutation tensor.
The first three terms are a bulk free energy density whose overall scale
is set by $A_0$ (discussed further below); $\gamma$ is a control parameter
related to a reduced temperature. Varying the latter in the absence of
chiral terms ($q_0=0$) gives an isotropic-nematic transition at
$\gamma = 2.7$ with a mean-field spinodal instability at $\gamma = 3$.

The remaining two terms of the free energy density in Eq.~\ref{free} describe
distortions of the order parameter field. In theoretical work and when
describing a generic rather than a specific system, it is 
conventional~\cite{mermin,deGennes} to
assume that splay, bend and twist deformations of the director are equally
costly, that is, there is a single elastic constant $K$. The parameter $q_0$
is related to the helical pitch length $p$ via $q_0=2\pi/p$,
describing one full turn of the director in the cholesteric phase.

There are two dimensionless numbers, which are commonly referred to as
$\kappa$, the chirality, and  $\tau$, the reduced temperature
\cite{mermin} which can be used to characterise the system. In terms of
the parameters in the free energy density, these are:
\begin{eqnarray}\label{cntrl-param} 
\tau&=&\frac{27(1-\gamma/3)}{\gamma}\label{tau}\\
\kappa&=&\sqrt{\frac{108\ K\, q_0^2}{A_0\, \gamma}}\label{kappa}.
\end{eqnarray}
If the free energy density Eq.~\ref{free} is made dimensionless, $\tau$
appears as prefactor of the term  quadratic in $Q_{\alpha\beta}$,
whereas $\kappa$
quantifies the ratio between bulk and  gradient free energy terms. The
chirality and reduced temperature may be used to characterise an equilibrium
phase diagram of the blue phases in bulk cholesteric liquid crystals
\cite{mermin,henrichfield}.

The effect of an electric field can be modelled by including the following
term in the bulk free energy density:
\begin{equation}
-\frac{\epsilon_\mathrm{a}}{12\pi} E_{\alpha}Q_{\alpha\beta}E_{\beta},
\end{equation} 
where $E_{\alpha}$ are the components of the electric field, and 
$\epsilon_\mathrm{a}$ (here assumed positive) is the dielectric anisotropy
of the liquid
crystal. The strength of the electric field is quantified via
one further dimensionless number
\begin{equation}
{\cal E}^2 = \frac{27 \epsilon_\mathrm{a}}{32 \pi A_0 \gamma} E_\alpha E_\alpha.
\end{equation}
The quantity ${\cal E}$ is known as the reduced field strength.

\medskip
\noindent
\textbf{Surface free energy.}
In addition to the fluid free energy density $f(Q_{\alpha\beta})$, a surface
free energy density is also present to represent the energetic cost of
anchoring at a solid surface.
In the case of normal anchoring, the surface free energy
density (per unit area) is
\begin{equation}
f_s(Q_{\alpha\beta}, Q^0_{\alpha\beta})
= {\textstyle \frac{1}{2}}W(Q_{\alpha\beta} - Q^0_{\alpha\beta})^2,
\end{equation}
where $Q^0_{\alpha\beta}$ is the preferred order
parameter tensor at the solid surface, and $W$ is a constant determining
the strength of the anchoring. In the case of planar anchoring, a slightly
more complicated expression is required to allow for degeneracy
(see, e.g., \cite{fournier2005}).
The determination of $Q^0_{\alpha\beta}$ for normal and planar anchoring is
discussed below. For a colloidal particle of radius $R$, the strength of
the surface
anchoring compared which the bulk fluid elastic constant may be quantified
by the dimensionless parameter $WR/K$. For small values of this parameter,
the presence of a particle surface should have little impact on the local
fluid LC ordering.

\medskip
\noindent
\textbf{Dynamics of the order parameter.}
A framework for the dynamics of liquid crystals is provided by the 
Beris-Edwards model \cite{beris}, in which the time evolution of the
tensor order parameter obeys
\begin{equation}
\label{eom}
\left(\partial_t+ u_\nu \partial_\nu \right) Q_{\alpha\beta} - S_{\alpha\beta}
= \mathit{\Gamma} H_{\alpha\beta}.
\end{equation}
In the absence of flow, Eq.~\ref{eom} describes a relaxation towards
equilibrium on a timescale determined by a collective rotational diffusion 
constant $\mathit{\Gamma}$. This relaxation is driven by the molecular field
$H_{\alpha\beta}$, which is the functional derivative of the free energy
with respect to the order parameter \cite{beris}:
\begin{equation}
H_{\alpha\beta} = -\frac{\delta {\cal F}} {\delta Q_{\alpha\beta}} 
+ {\textstyle \frac{1}{3}} \delta_{\alpha\beta} 
\mathrm{Tr}\left(\frac{\delta {\cal F}}{\delta Q_{\alpha\beta}}\right).
\end{equation}

The tensor $S_{\alpha\beta}$ in Eq.~\ref{eom} completes the material
derivative for rod-like molecules \cite{beris}. It couples the order
parameter to the symmetric and antisymmetric parts of the velocity 
gradient tensor $W_{\alpha \beta}\equiv\partial_\beta u_\alpha$.
The symmetric part $A_{\alpha\beta}$ and the antisymmetric part
$\Omega_{\alpha\beta}$ are defined as 
\begin{equation}
A_{\alpha\beta} = {\textstyle \frac{1}{2}} (W_{\alpha\beta} + W_{\beta\alpha}),
\end{equation}
and
\begin{equation}
\Omega_{\alpha\beta} = {\textstyle \frac{1}{2}} (W_{\alpha\beta} - W_{\beta\alpha}).
\end{equation}
This full coupling term is then
\begin{eqnarray}
\label{coupling-term}
S_{\alpha\beta}  &=&
(\xi A_{\alpha\nu} + \Omega_{\alpha\nu})
(Q_{\nu\beta} + {\textstyle \frac{1}{3}}\delta_{\nu\beta})
+
(Q_{\alpha\nu} + {\textstyle \frac{1}{3}} \delta_{\alpha\nu})
(\xi A_{\nu\beta} - \Omega_{\nu\beta})
\nonumber\\ 
&-&2 \xi ({Q_{\alpha\beta} + {\textstyle \frac{1}{3}}\delta_{\alpha\beta}})
Q_{\nu\nu} W_{\nu\nu}.
\end{eqnarray}
Here, $\xi$ is a material-dependent that controls the
relative importance of rotational and elongational flow for molecular
alignment, and determines in practice how the orientational order
responds to a local shear flow. 
In all cases, the value used here is $\xi = 0.7$, which is within
the `flow aligning' regime, where molecules align at a fixed angle
(the Leslie angle) to the flow direction in weak simple shear \cite{deGennes}.
The value of the collective rotational diffusion used in all simulations is
$\mathit{\Gamma} = 0.5$ in simulation units.

\medskip
\noindent
\textbf{Hydrodynamics.}
The momentum evolution obeys a Navier-Stokes equation driven by the divergence
of a generalised stress $P_{\alpha\beta}$:
\begin{equation}
\label{nse}
\rho\,\partial_tu_\alpha +\rho \,u_\beta \partial_\beta u_\alpha
=\partial_\beta P_{\alpha\beta}.
\end{equation}
The pressure tensor $P_{\alpha\beta}$ is, in general, asymmetric and includes
both viscous and thermodynamic components:
\begin{eqnarray}
P_{\alpha \beta}&=&
- p_0 \delta_{\alpha \beta} 
+ \eta \{ \partial_\alpha u_\beta + \partial_\beta u_\alpha\}
\nonumber\\
&-&  \xi H_{\alpha \gamma}\left(Q_{\gamma \beta}
+ {\textstyle \frac{1}{3}} \delta_{\gamma \beta} \right)
-\xi \left(Q_{\alpha \gamma}
+ { \textstyle \frac{1}{3}} \delta_{\alpha \gamma}\right) H_{\gamma \beta}
\nonumber\\ 
&+& 2 \xi  \left(Q_{\alpha \beta}
+ {\textstyle \frac{1}{3}} \delta_{\alpha \beta}\right) Q_{\gamma \nu} H_{\gamma \nu}
-\partial_\alpha Q_{\gamma \nu}
\frac{\delta{\cal F}}{\delta \partial_{\beta} Q_{\gamma \nu}}\nonumber\\
&+& Q_{\alpha \gamma}H_{\gamma \beta}-H_{\alpha \gamma} Q_{\gamma \beta}.
\end{eqnarray}
A lattice Boltzmann (LB) flow solver is used, where the isotropic pressure
$p_0$
and viscous terms are managed directly by the solver (as in a simple Newtonian
fluid, of viscosity $\eta$), whereas the divergence of the remaining terms
is treated as a local force density on the fluid. In all simulations
reported the mean fluid density $\rho = 1$ and the viscosity
$\eta = 0.01$ (simulations in confined geometry also used enhanced
bulk viscosity $\zeta = 10\eta$ to ensure numerical stability).

\medskip
\noindent
\textbf{Surface boundary conditions for particles.}
Hydrodynamic boundary conditions for solid objects are handled via
the lattice Boltzmann solver. In particular, the standard method of
bounce-back on links is used for both particles and confining walls
\cite{ladd94,nguyen2002}. Boundary conditions for the order parameter
are set out in \cite{juho}.

In all the simulations reported here the preferred orientation of the
director field at the particle surface is normal to the surface. The
required nematic director at the particle surface in a given location
$\hat{n}^0_\alpha$ may then be determined from geometry alone,
and a preferred order parameter $Q_{\alpha\beta}^0$ is computed via
\begin{equation}
Q^0_{\alpha\beta} = q^0(\hat{n}_\alpha^0 \hat{n}_\beta^0 
- {\textstyle \frac{1}{3}} \delta_{\alpha\beta}).
\label{eq:surf_q}
\end{equation}
The magnitude of surface order $q^0$ is set by
\begin{equation}
q^0 = {\textstyle \frac{2}{3}} \left( {\textstyle \frac{1}{4}} 
+ {\textstyle \frac{3}{4}} \sqrt{1 + (8/3\gamma)} \right)
\label{eq:surf_q0}
\end{equation}
which corresponds to the bulk order in a defect-free nematically ordered 
sample~\cite{denniston} (this is also very close to the value of the order 
parameter in a cholesteric or in a blue phase away from disclinations).

In all simulations the colloidal (hard sphere) radius is $R = 2.3$. To
prevent particles overlapping at the hard sphere radius, an additional
short range soft potential is included. This takes the form of
$V(h) = \epsilon (\sigma/h)$ where $h$ is the surface to surface separation.
The parameters are $\epsilon = 0.0004$ and $\sigma = 0.1$ in simulation
units. Both potential and resulting force are smoothly matched to zero at
a cut-off distance of $h_\mathrm{c} = 0.25$ simulation units.

\medskip
\noindent
\textbf{Surface boundary conditions for walls.}
The confining solid walls, which are flat and stationary, have the
same normal boundary condition as at the particle surface. In addition,
degenerate planar anchoring is applied at the walls, where the
preferred nematic director may adopt any orientation parallel
to the surface. The preferred direction is determined explicitly
by projecting the local fluid order parameter to the plane of the
wall \cite{fournier2005}. Eq.~\ref{eq:surf_q} and Eq.~\ref{eq:surf_q0}
are then used as before.

To prevent the particles being forced into the wall, a correction
to the lubrication force on the colloidal particle is added for very low
particle
surface to wall surface separations $h < 0.5$ lattice spacing. The
correction is based on the analytic expression for the lubrication
between a sphere and a flat surface.

\pagebreak

\noindent
\textbf{\large Acknowledgements}

\noindent
We acknowledge support by UK EPSRC grants EP/J007404/1 (KS and JSL), 
EP/G036136/1 (OH), and EP/I030298/1 (KS and JSL).
MEC is supported by the Royal Society.
We thank both Hector and PRACE for computational resources.

\bigskip
\noindent
\textbf{\large Author Contributions}

\noindent
KS, OH and JSL contributed to the code base for simulation and performed
simulations. All authors helped to analyse the results and write the
manuscript.

\bigskip
\noindent
\textbf{\large Additional Information}

\noindent
\textbf{Competing financial interests:} The authors declare no competing
financial interests.

\newpage

\begin{figure}[p!]

\centerline{\includegraphics[width=\textwidth]{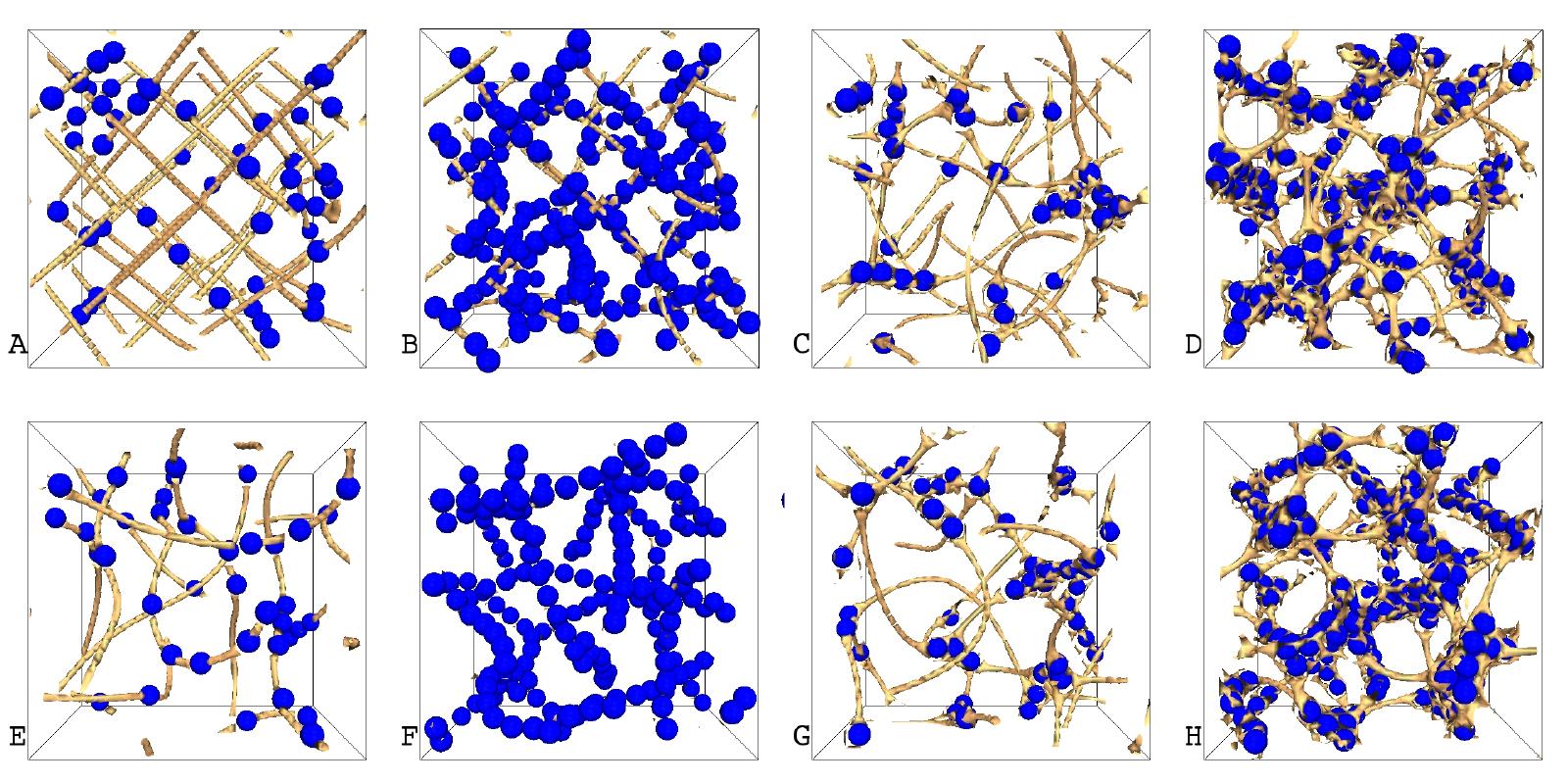}}
\caption{\textbf{Bulk blue phase structures.} Snapshots of the states
obtained after dispersing
a suspension of colloidal nanoparticles within a cholesteric liquid
crystal in the BPI-forming region. (\textbf{a})--(\textbf{d}) correspond
to the case
of a dispersion of particles in a pre-equilibrated BPI phase;
(\textbf{e})--(\textbf{h}) are obtained by dispersing colloids in an isotropic
phase and then quenching into the range where BPI is stable, leading to
formation of an amorphous, BPIII-like, disclination network.
Structures correspond to:
(\textbf{a}) and (\textbf{e}) $w=0.23$ and $\phi=1\%$;
(\textbf{b}) and (\textbf{f}) $w=0.23$ and $\phi=5\%$; 
(\textbf{c}) and (\textbf{g}) $w=2.3$ and $\phi=1\%$;
(\textbf{d}) and (\textbf{h}) $w=2.3$ and $\phi=5\%$.
The anchoring of the director field to the colloidal surface is normal.
[For the full parameter list used to generate Figs.~1-3 see Supplementary
Notes.] For clarity, only a portion of the simulation box is shown
its linear extent being $\simeq 28R$
(one eighth of the total volume);
the full structures are shown in
Supplementary Figs.~2 and~3.}
\end{figure}

\begin{figure}[p!]
\centerline{\includegraphics[width=\textwidth]{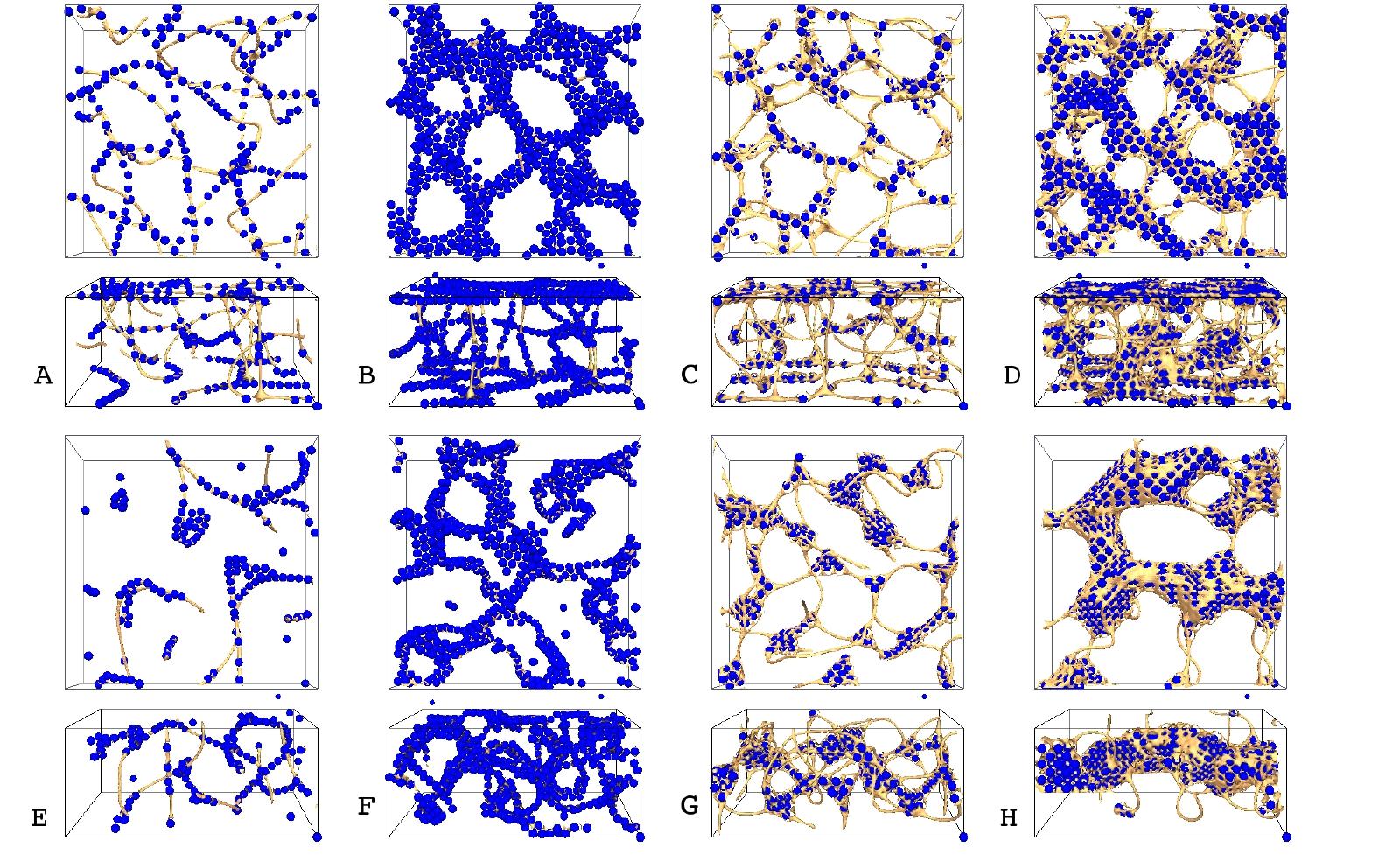}}
\caption{\textbf{Confined blue phase structures.}
Snapshots of the steady states obtained when a dispersion of colloids in
the isotropic phase is placed in a sandwich geometry, 
and then quenched into a regime where BPI is stable in the bulk.
Both top and side views are provided.
The anchoring of the director field at the walls is normal for
(\textbf{a})--(\textbf{d}) and planar for (\textbf{e})--(\textbf{h}). 
Structures correspond to:
(\textbf{a}) and (\textbf{e}) $w=0.23$ and $\phi=1\%$; 
(\textbf{b}) and (\textbf{f}) $w=0.23$ and $\phi=2\%$; 
(\textbf{c}) and (\textbf{g}) $w=2.3$ and $\phi=1\%$;
(\textbf{d}) and (\textbf{h}) $w=2.3$ and $\phi=2\%$.
As in Fig.~1, only one quarter of the simulation box is shown for clarity
(its horizontal extent being $\simeq 56R$ and vertical
extent $\simeq 24R$ in the narrow direction);
the full structures are shown in Supplementary Figs.~5 and~6.}
\end{figure}

\begin{figure}[p!]
\begin{center}
\includegraphics[width=0.48\textwidth]{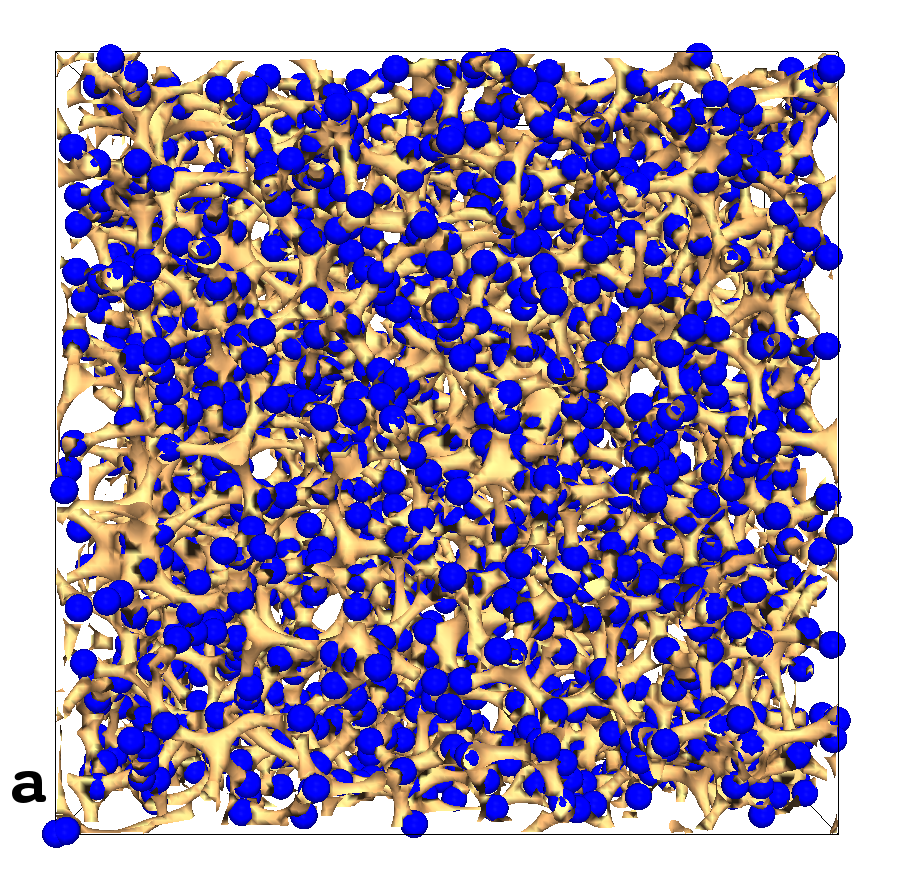}
\includegraphics[width=0.43\textwidth]{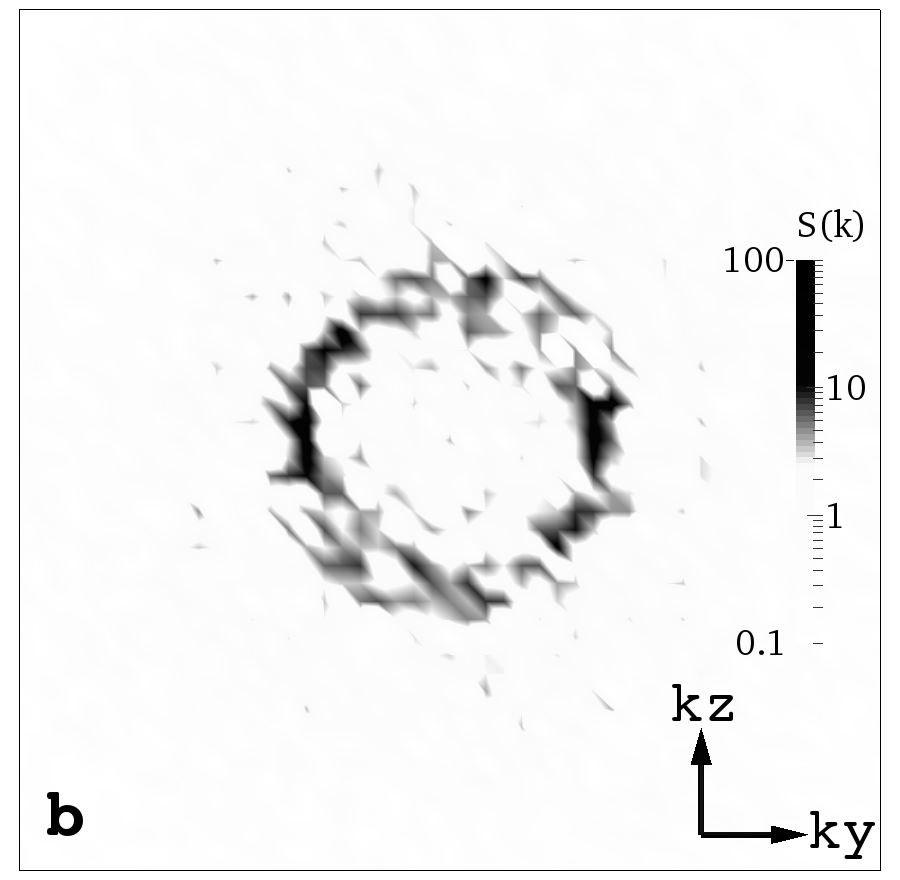}\\
\includegraphics[width=0.48\textwidth]{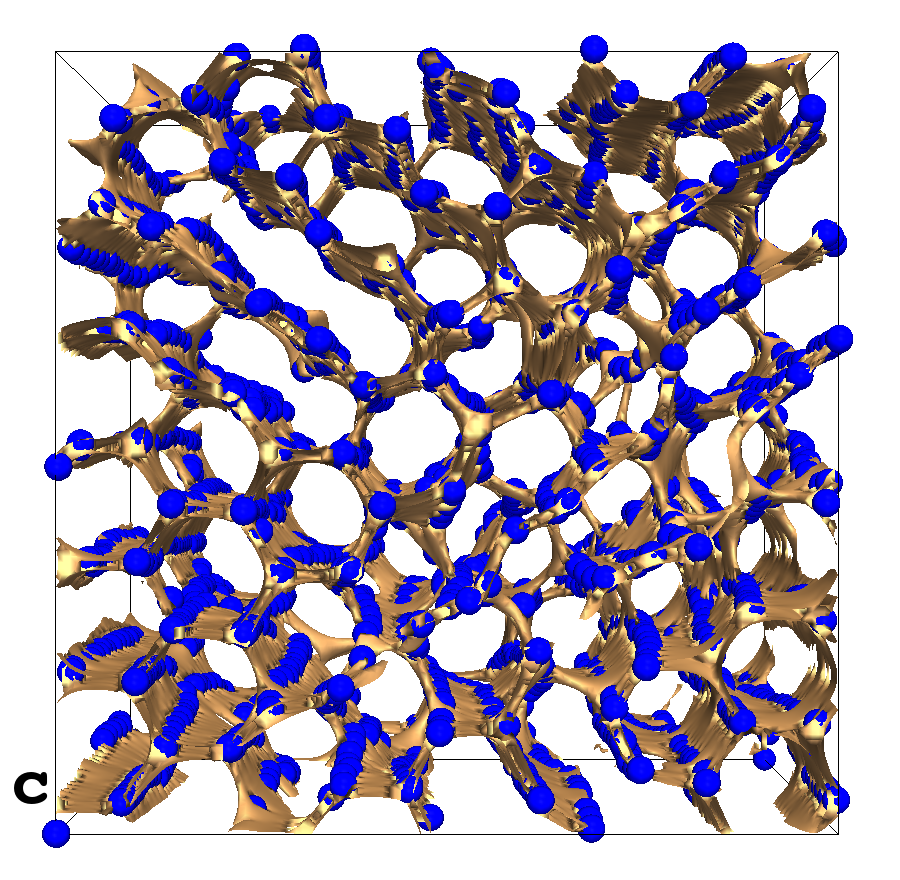}
\includegraphics[width=0.43\textwidth]{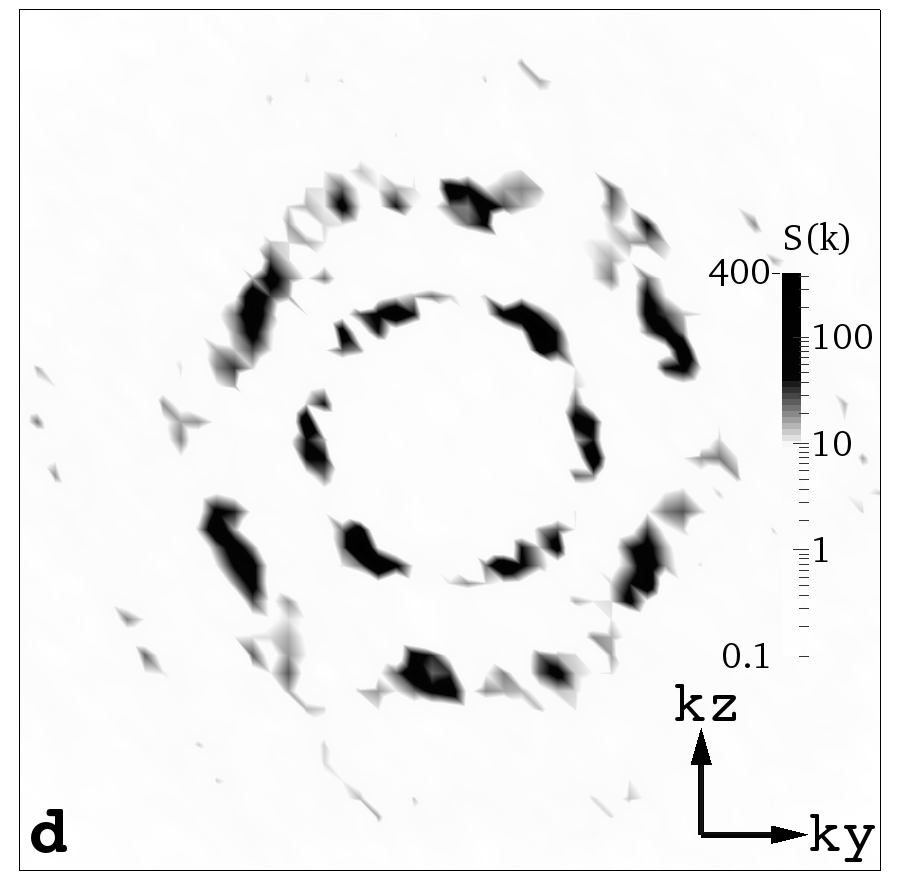}
\end{center}
\caption{\textbf{Structure in electric field.}
Steady states obtained when a dispersion of colloids is inserted
into a thermodynamically stable bulk BPIII with and without applied
external electric field. Panel~(\textbf{a}) shows the situation in
zero field, and panel~(\textbf{c})
the situation when the applied field is in the $x$-direction
(perpendicular to the plane of the paper). Panels (\textbf{b})
and~(\textbf{d}) show one
plane of the three-dimensional structure factor $S(\mathbf{k})$ computed
from the positions of the particles, with wavevector $k_x = 0$, corresponding
to~(\textbf{a}) and~(\textbf{c}) respectively. With field, the disclinations
form a honeycomb
pattern with hexagonal ordering perpendicular to the field direction;
this is reflected in the structure factor.
Simulations use weakly anchoring particles $w=0.23$, and a volume
fraction of $\phi=5\%$ (see Supplementary Notes for full list of
parameters).
The full simulation with a box size of $\simeq 56R$ is shown.}
\end{figure}

\newpage
Blank page.
\newpage

\newpage
Blank page.
\newpage

{\bf Supplementary Figures}

\begin{figure}[!h]
\begin{center}
\includegraphics[scale=0.8]{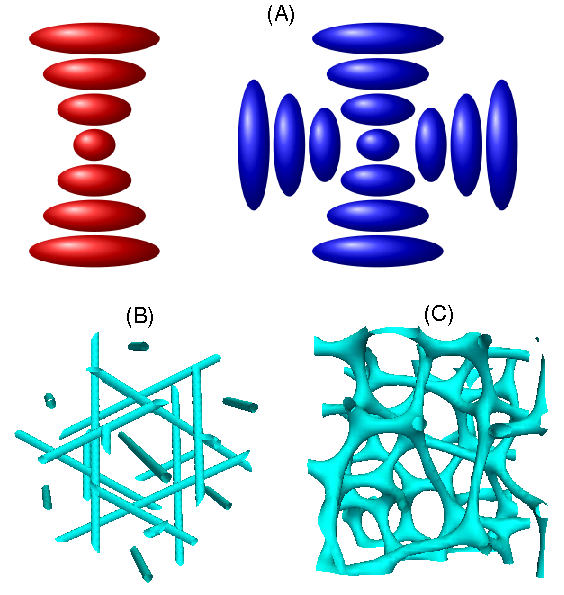}
\end{center}
\caption{\textbf{Blue phase structures.}
(A) Schematic representation of the director field
in a cholesteric (left) and within a double twist cylinder (right).
(B) Snapshot of the disclination network in an equilibrated 
blue phase I structure. (C) Same as (B) for an amorphous blue phase III
network.}
\end{figure}

\newpage

\begin{figure}[!h]
\begin{center}
\includegraphics[scale=0.35]{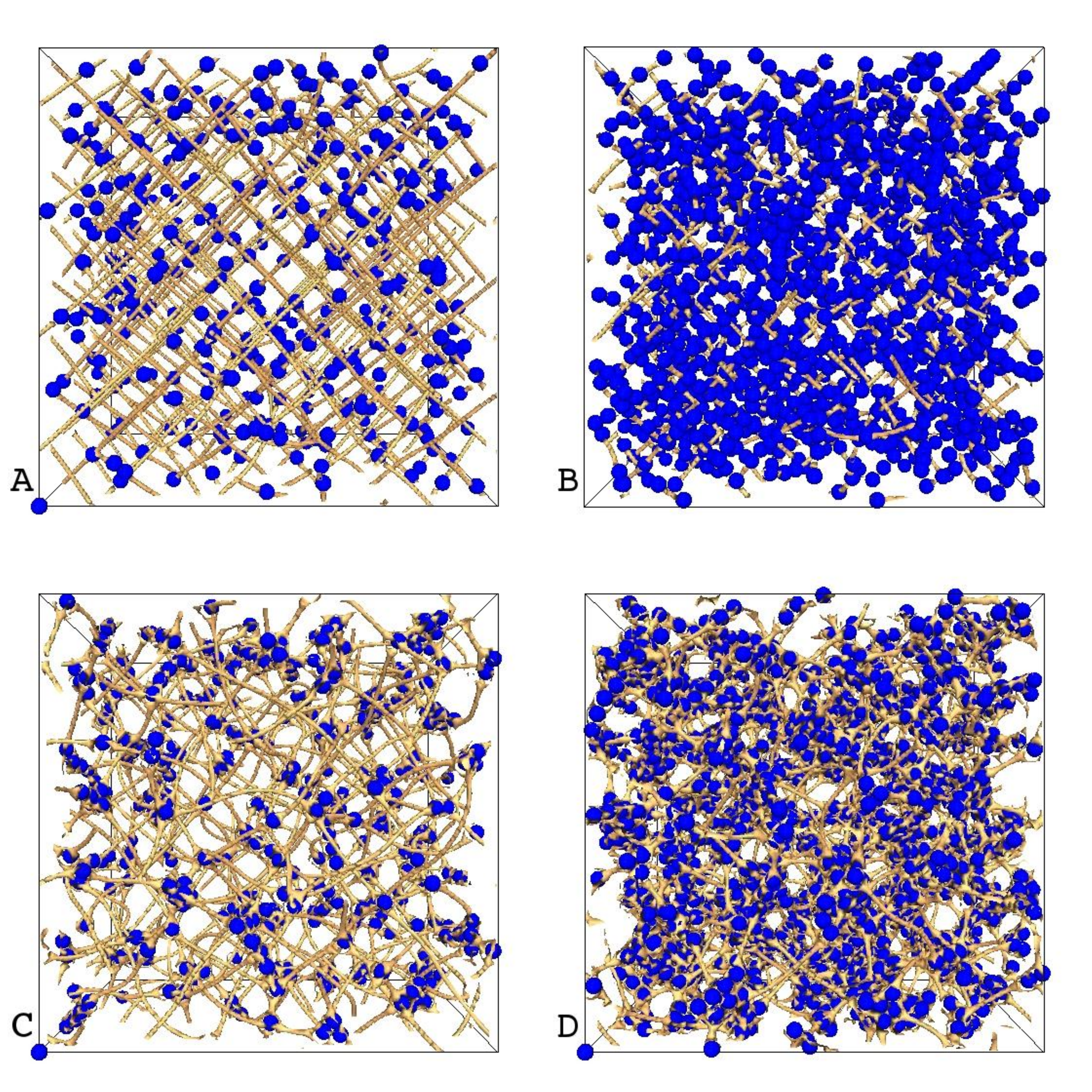}
\end{center}
\caption{\textbf{Bulk blue phase structures.}
As for main manuscript Fig.~1~\textbf{a}--\textbf{d}
(with liquid crystal order parameter
initialised to equilibrium blue phase~I structure), but for the entire
simulation system of 128$^3$ lattice sites. (A)~solid volume fraction~1\%
and weak anchoring;  (B)~4\% solid volume fraction and weak anchoring;
(C) 1\% solid volume fraction and strong anchoring; (D) 4\% solid
volume fraction and strong anchoring. The view direction in the
main figure is from the left here. There is a reference
particle at the bottom left in each panel which does not take part
in the simulation.}
\end{figure}

\newpage

\begin{figure}[!h]
\begin{center}
\includegraphics[scale=0.35]{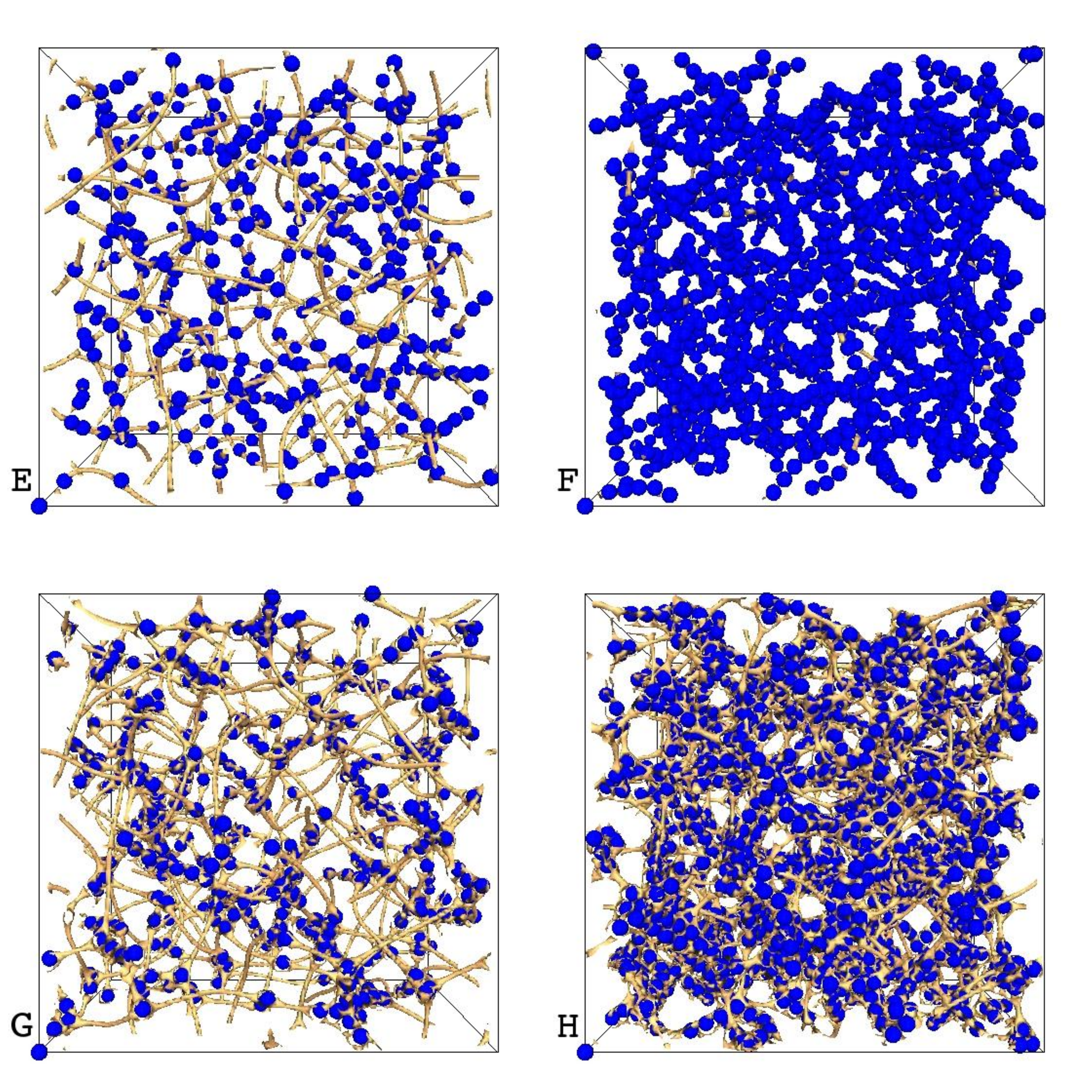}
\end{center}
\caption{\textbf{Bulk blue phase structures.}
As for main manuscript Fig.~1 \textbf{e}--\textbf{h}
(initialised via a ``quench''
to generate a disordered network), but for the entire simulation
system of 128$^3$ lattice sites. (E) solid volume fraction 1\% and
weak anchoring; (F) 4\% solid volume fraction and weak anchoring;
(G) 1\% solid volume fraction and strong anchoring; (H) 4\% solid
volume fraction and strong anchoring. Again, there is a reference
particle at the bottom left in each case which does not take part
in the simulation.}
\end{figure}

\newpage

\begin{figure}[!h]
\begin{center}
\includegraphics[scale=0.49]{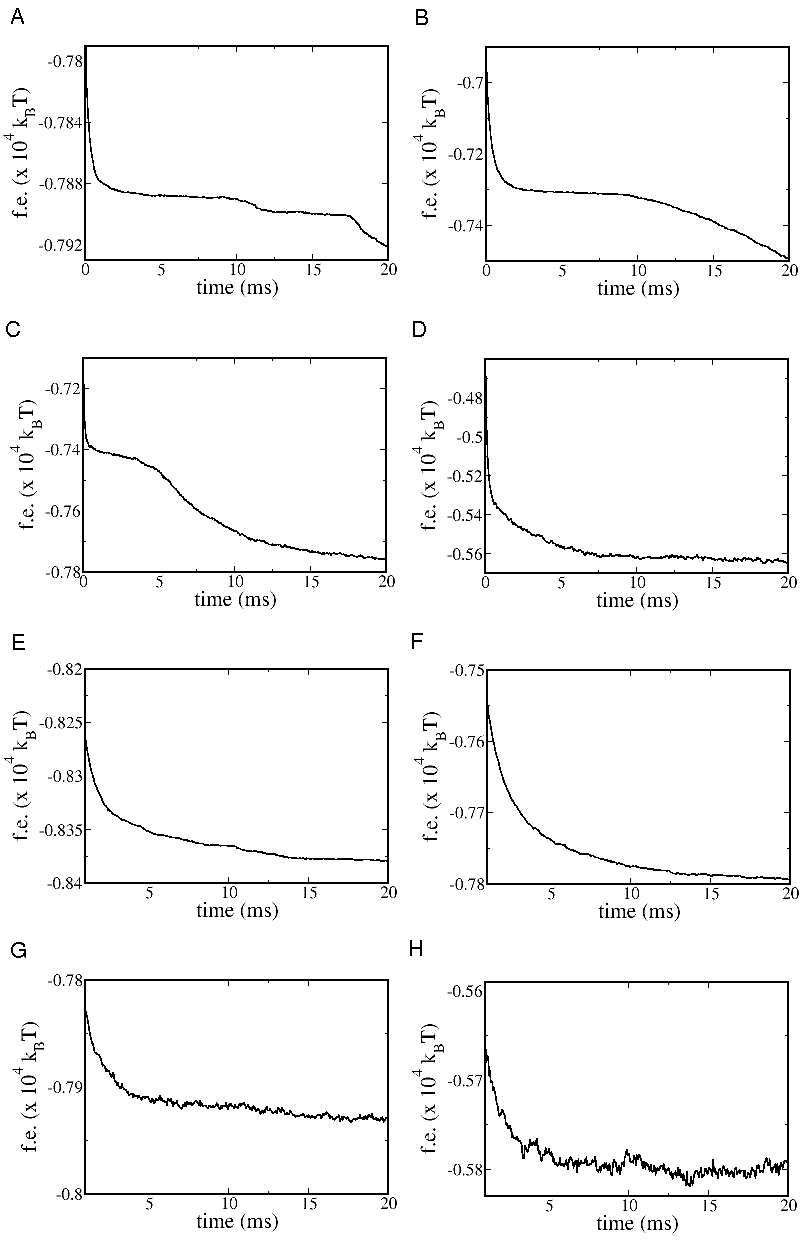}
\end{center}
\caption{\textbf{Free energy evolution.}
Plot of the free energy (per BP unit cell) versus time
for the simulations corresponding to Fig.~1 (\textbf{a}--\textbf{h})
in the main text.}
\end{figure}

\newpage

\begin{figure}[!h]
\begin{center}
\includegraphics[scale=0.42]{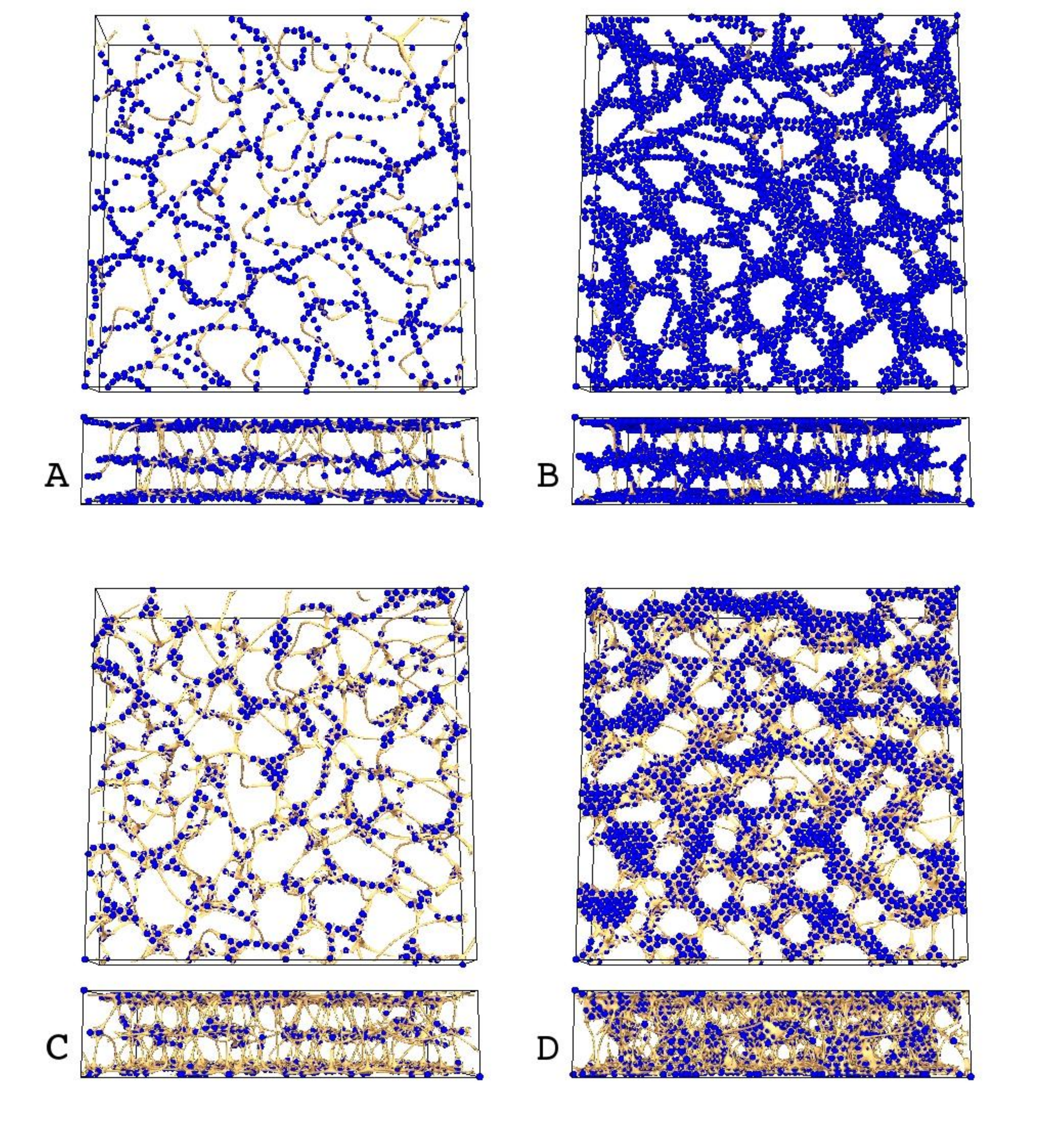}
\end{center}
\caption{\textbf{Confined blue phase structures.}
As for main manuscript Fig.~2 \textbf{a}--\textbf{d}
(confined geometry with normal
anchoring walls at both top and bottom), but for the entire simulation
system of 256$^2\times$56 lattice sites. Each shows a top view and
side view of the same simulation state. (A) solid volume fraction 1\%
and weak anchoring; (B) 4\% solid volume fraction and weak anchoring;
(C) 1\% solid volume fraction and strong anchoring; (D) 4\% solid
volume fraction and strong anchoring.}
\end{figure}

\newpage

\begin{figure}[!h]
\begin{center}
\includegraphics[scale=0.42]{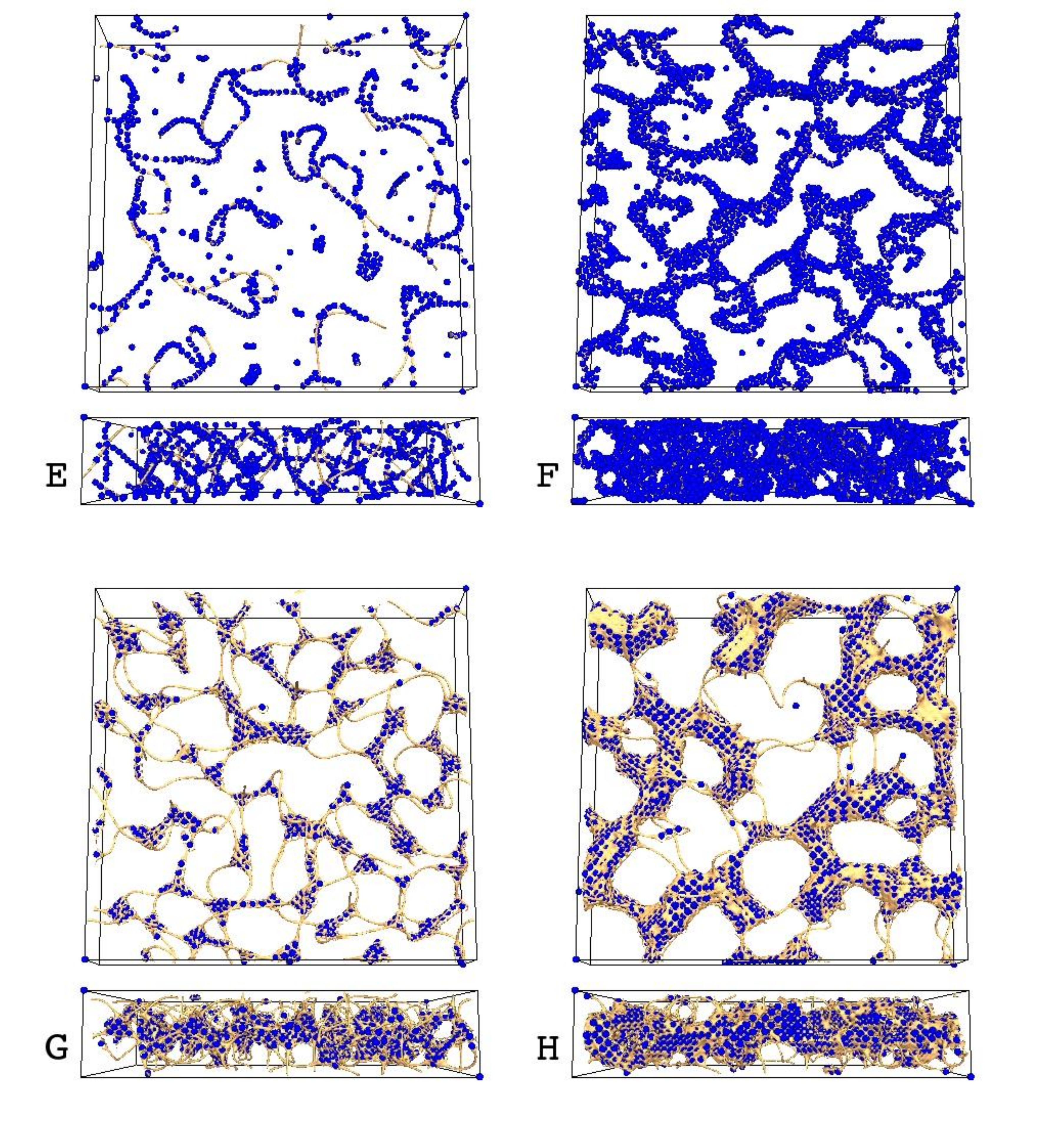}
\end{center}
\caption{\textbf{Confined blue phase structures.}
As for main manuscript Fig.~2 \textbf{e}--\textbf{h}
(confined geometry with planar
anchoring walls at both top and bottom), but for the entire simulation
system of 256$^2\times$56 lattice sites. Each shows a top view and
side view of the same simulation state. (E) colloid solid volume fraction 1\%
and weak anchoring; (F) 4\% solid volume fraction and weak anchoring;
(G) 1\% solid volume fraction and strong anchoring; (H) 4\% solid
volume fraction and strong anchoring.}
\end{figure}

\newpage

\begin{figure}[!h]
\begin{center}
\includegraphics[width=0.34\columnwidth]{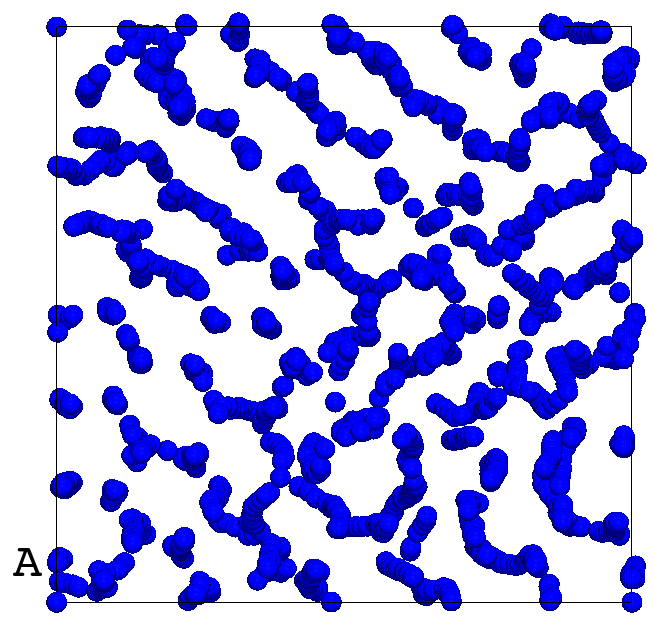}
\includegraphics[width=0.32\columnwidth]{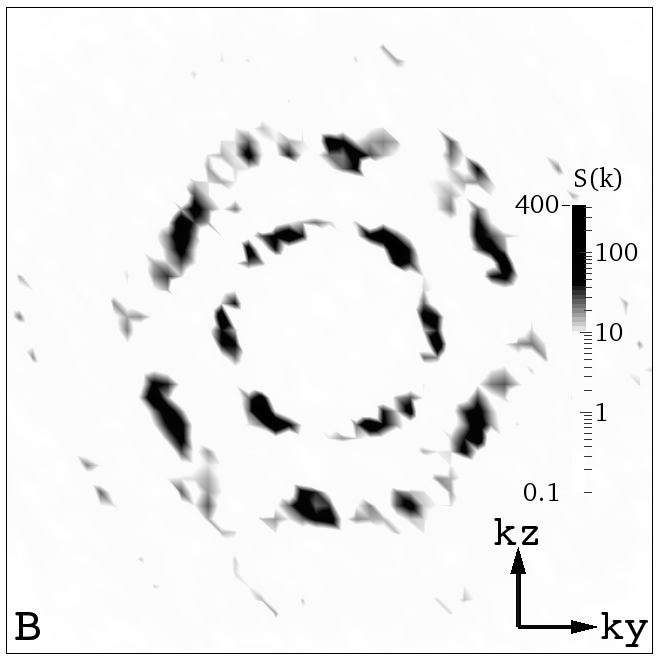}
\includegraphics[width=0.32\columnwidth]{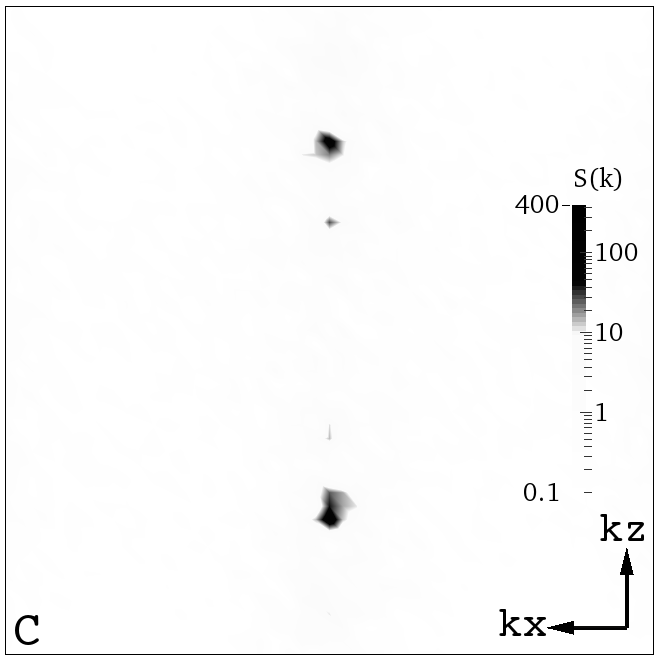}
\includegraphics[width=0.34\columnwidth]{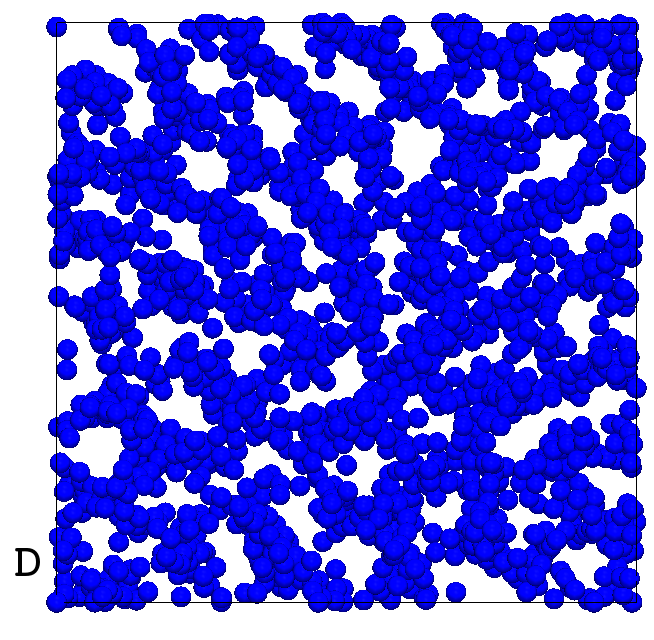}
\includegraphics[width=0.32\columnwidth]{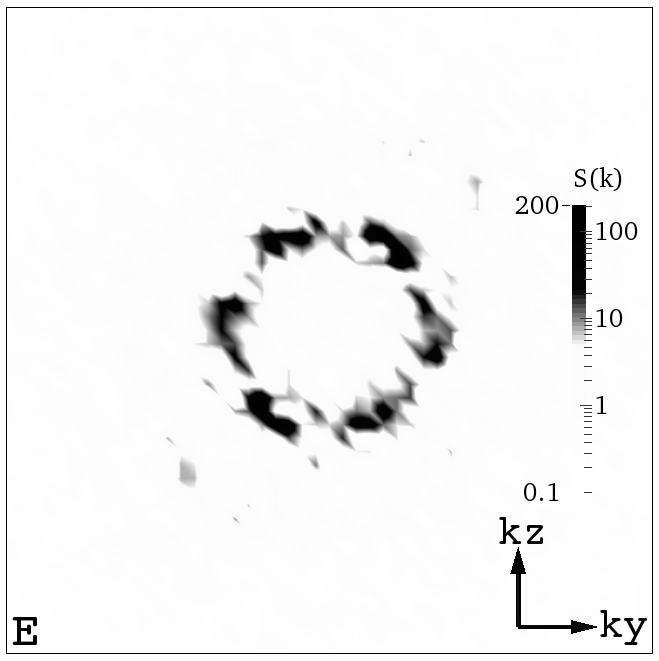}
\includegraphics[width=0.32\columnwidth]{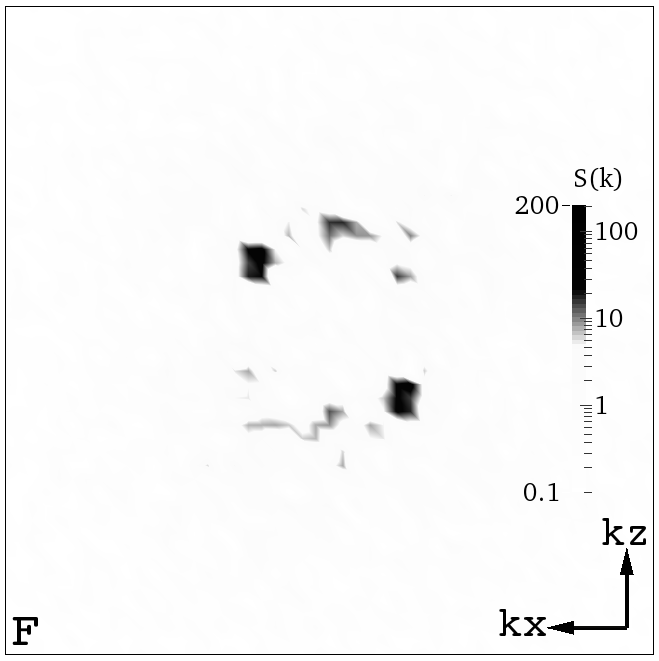}
\end{center}
\caption{\textbf{Structure with and without electric field.}
(Disclination lines are 
removed for clarity.) Panels~A and~B reproduce the results in Fig.~3\textbf{c}
and
Fig.~3\textbf{d} respectively with the applied field in the $x$-direction 
(out of the plane of the paper). Also included is a cut of the structure 
factor with $k_y = 0$ (panel C). 
Panels D, E, and F show the corresponding situation after the
external field has been removed and the structure allowed to relax,
and show residual hexagonal ordering. The structure factor data is for wave
vectors $k_x = 0; (k_y, k_z) \in [-3\pi/8 , 3\pi/8]$ (panels B and E) and
$k_y = 0; (k_x, k_z) \in [-3\pi/8, 3\pi/8]$ (panels C and F).}
\end{figure}

\newpage

\begin{figure}[!h]
\begin{center}
\includegraphics[width=0.34\columnwidth]{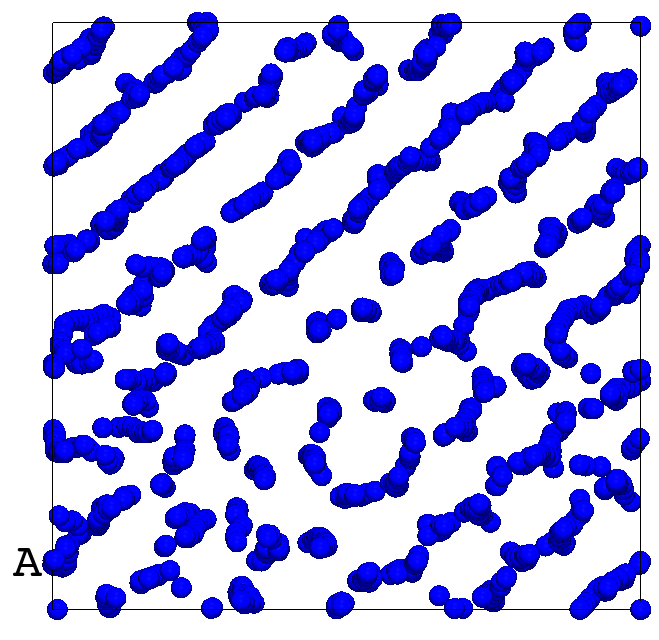}
\includegraphics[width=0.32\columnwidth]{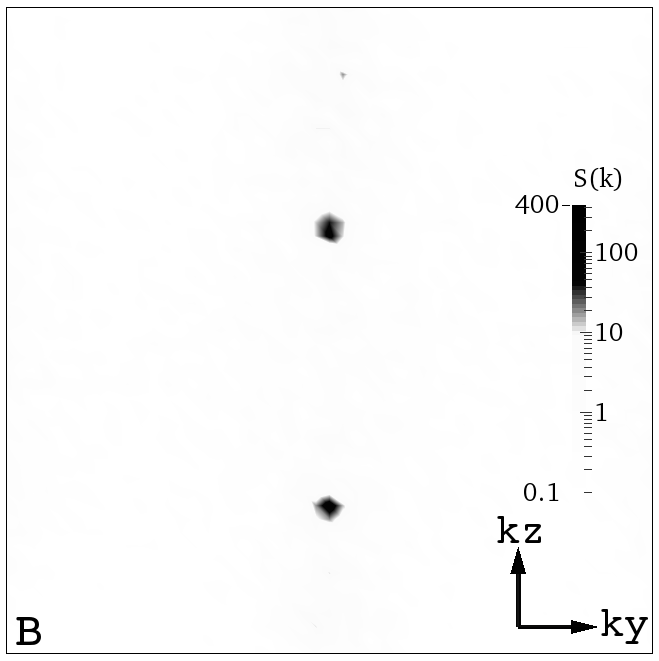}
\includegraphics[width=0.32\columnwidth]{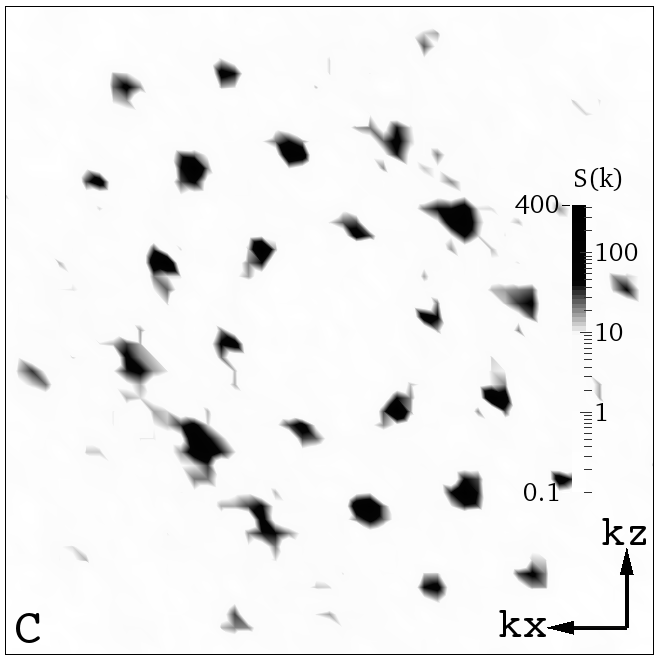}\\
\includegraphics[width=0.34\columnwidth]{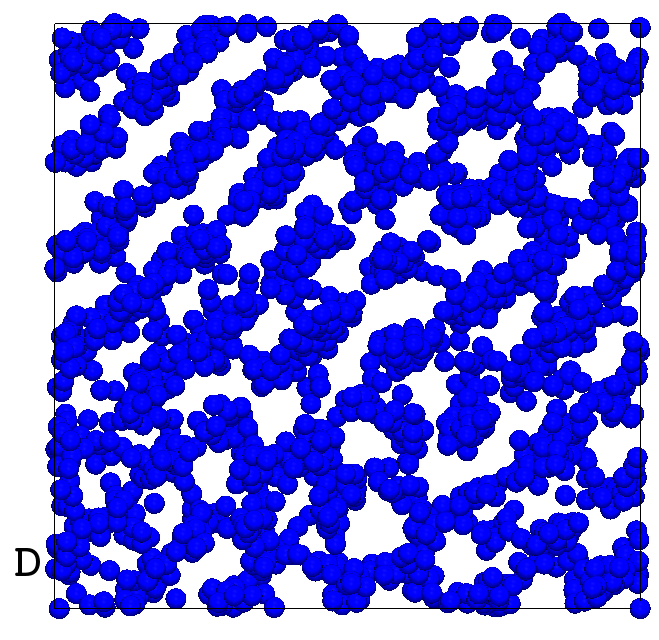}
\includegraphics[width=0.32\columnwidth]{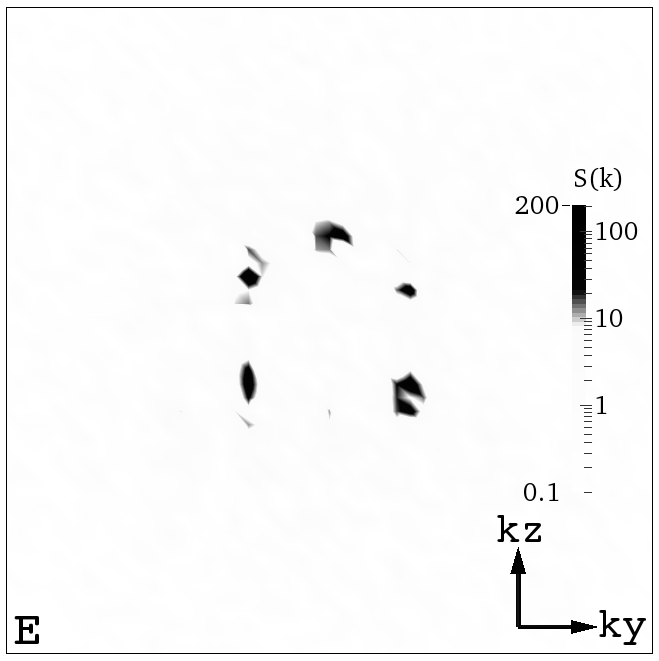}
\includegraphics[width=0.32\columnwidth]{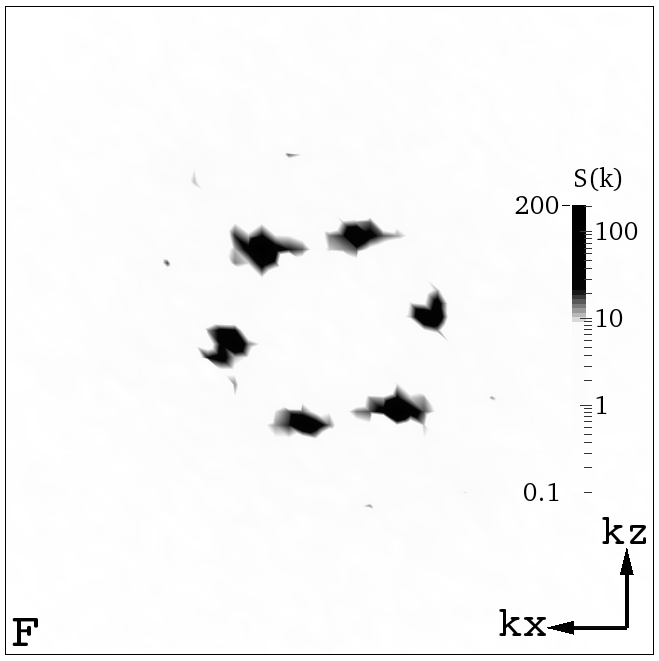}\\
\end{center}
\caption{\textbf{Cycling the electric field.}
Starting from the final position of
Supplementary Fig.~7 panel~D, the field is re-applied, this time in the
$y$-direction. The corresponding view of the particle distribution is
shown, again viewing along the field direction, in panel A.
Corresponding structure factor cuts with $k_x = 0$ and $k_y = 0$ are
shown in panels~B and~C, respectively. The corresponding situation when
the field is again switched off, and the structure allowed to relax,
is shown in panels D, E, and~F. The wave vector values used are the
same as in Supplementary Fig.~7.}
\end{figure}

\newpage

\begin{figure}[!h]
\begin{center}
\includegraphics[width=0.34\columnwidth]{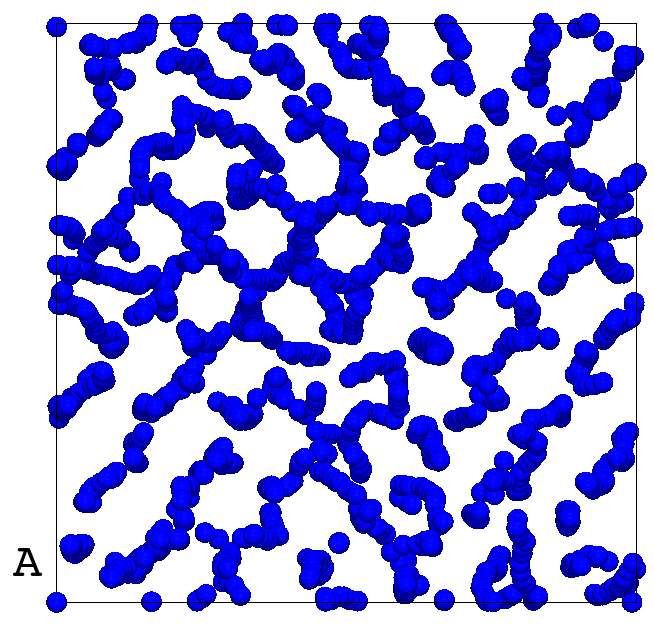}
\includegraphics[width=0.32\columnwidth]{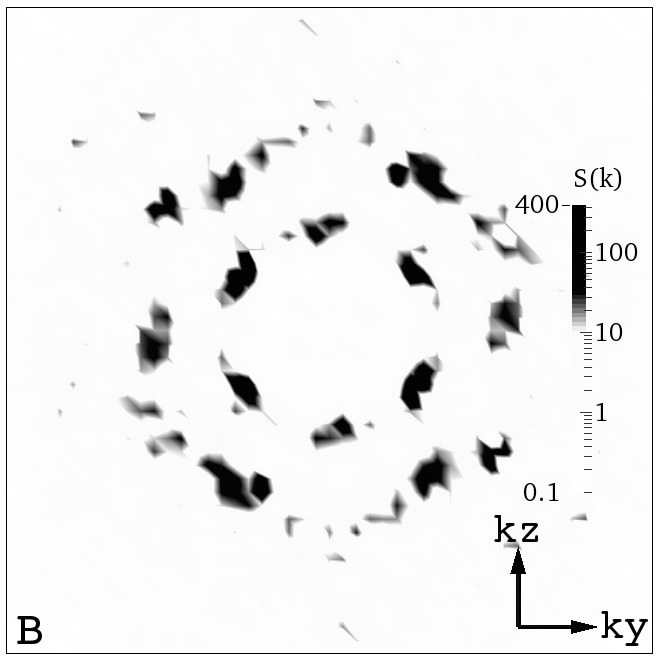}
\includegraphics[width=0.32\columnwidth]{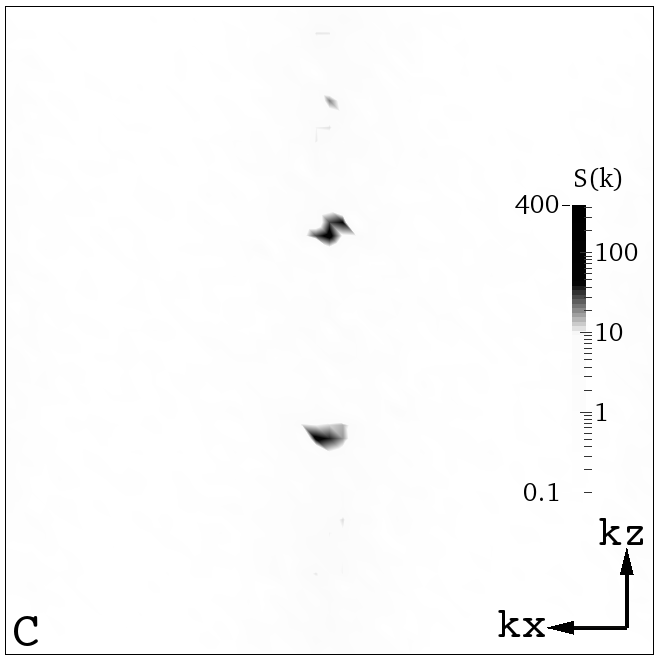}\\
\includegraphics[width=0.34\columnwidth]{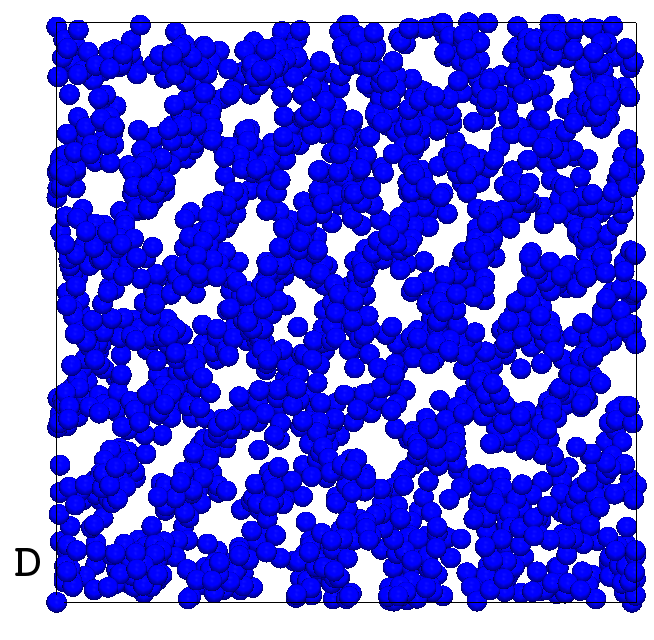}
\includegraphics[width=0.32\columnwidth]{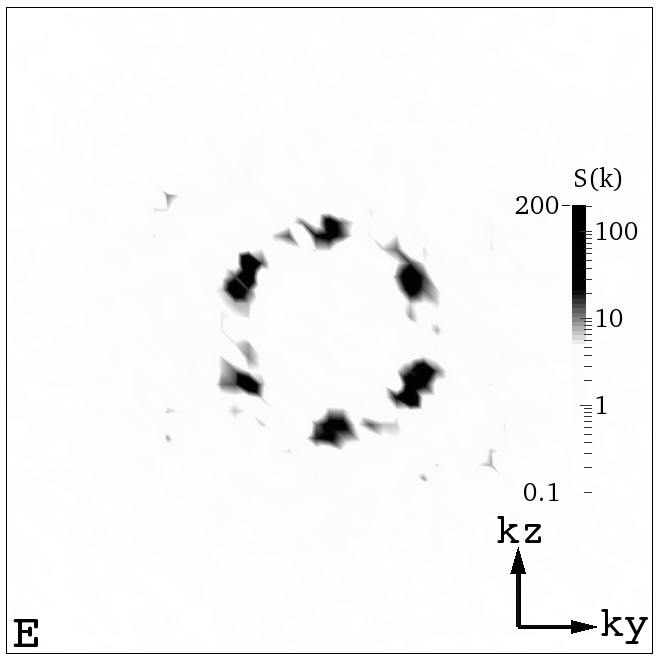}
\includegraphics[width=0.32\columnwidth]{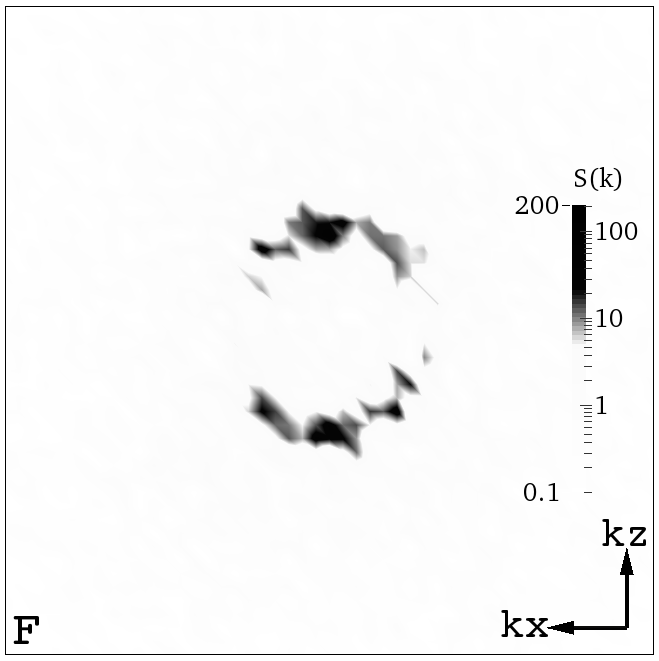}
\end{center}
\caption{\textbf{Cycling the electric field further.}
Starting from the final
position of Supplementary Fig.~8 panel~D, the field is re-applied
again in the $x-$direction. Panel A shows the particle configuration
viewed along the field direction, and panels B and C show the structure
factor with $k_x = 0$ and $k_y = 0$ respectively. The corresponding
situation when the field is removed and the structure allowed to relax
is shown in panels D, E, and F. The wave vector values are again as
in Supplementary Fig.~7. Note that the on and off states are 
equivalent to those in Supplementary Fig.~7, up to a rotation.
This is because all hexagonal structures perpendicular to the field
direction are energetically degenerate,
and the actual orientation of the hexagon emerges spontaneously during
switching. }
\end{figure}

\vfill\pagebreak

{\bf Supplementary Table}

\begin{table}[h]
\begin{center}
\begin{tabular}{|l|c|c|c|c|c|c|c|}
\hline
         & $A_0$ & $\gamma$ & $K$ & $p$ & $\kappa$ & $\tau$ & $WR/K$\\
\hline 
Bulk BPI & 0.01 & 3.086 & 0.007061 & 32$\sqrt{2}$ & 0.6902 & -0.2500 &
0.23, 2.3\\
\hline
Bulk Quench & 0.01 & 3.086 & 0.007061 & 32$\sqrt{2}$ & 0.6902 & -0.2500 &
0.23, 2.3 \\
\hline
Confined & 0.01 & 3.086 & 0.01897 & 64 & 0.8 & -0.2500 & 0.23, 2.3\\
\hline
External field & 0.004& 3.086 & 0.02 & 32 & 2.6 & -0.2500 & 0.23\\ 
\hline
\end{tabular}
\end{center}
\caption{\textbf{Free energy parameters used in the simulations.} Bulk BPI
correspond to Fig.~1 (\textbf{a}--\textbf{d}), bulk quench to
Fig.~1 (\textbf{e}--\textbf{h}); confined
geometries are shown in Fig.~2; and the external field case is
relevant for Fig.~3.}
\label{table:params}
\end{table}

\vfill
\pagebreak

{\bf Supplementary Methods}

\noindent
\textbf{Parameter details for bulk BPI.}
The parameters for the free energy are chosen to be representative
of blue phase I: chirality $\kappa \sim 0.7$ and reduced temperature
$\tau = -0.25$.
The full free energy parameters are shown in Table~\ref{table:params}.
The order parameter tensor $Q_{\alpha\beta}$ for bulk blue phase I is
initialised from an approximation in high chirality limit
\cite{supportblue1,support-oliver1}. Systems of 128$^3$ lattice sites are
used with a pitch length of $p = 32\sqrt{2}$, which accommodates 4$^3$
unit cells with fully periodic boundary conditions.
Simulations at different solid volume fractions (1\% and 4\%), and different 
surface anchoring strengths ($WR/K = 0.23$ and $WR/K = 2.3$) representing
``weak'' and ``strong'' anchoring are run for two million simulation
time steps.
The initial distribution of the colloid positions is set at random
for the required solid volume fraction; the colloids are initially
at rest.

To generate a disordered network of disclination lines, a ``quench''
is performed in which the order parameter is initialised via a
locally chosen random director, and
Eq.~15
employed with a small amplitude $q^0 = 10^{-7}$ to
provide an order parameter.
The mean-field spinodal point at $\gamma = 3.0$ is avoided and
the parameters are the same as for the initial simulations with
pitch $p = 32\sqrt{2}$. The chirality and reduced temperature (see
Table~\ref{table:params}) remain appropriate for BPI.
Colloids are added as before and the simulations run for 2 million
simulation steps.

For simplicity,  the size of the BPI unit cell is not allowed to 
readjust dynamically to minimise free energy (there is no 
``redshift''~\cite{supportblue1}): this would lead to quantitatively different 
values of the free energy, but qualitatively similar structures. 

\noindent
\textbf{Parameters for confined BPI.}
Here, a narrow sandwich of fluid is placed between flat walls in
perpendicular to the narrow coordinate direction, and with periodic
boundaries in the other two directions. The system size used in all
cases is 256$^2 \times$56 lattice sites. Each simulation is a ``quench''
in a similar manner to that used in the bulk: the order parameter is
initialised randomly. The free energy parameters (see Table~\ref{table:params})
are again appropriate for (equilibrium) blue phase I
(chirality $\kappa = 0.8$ and
reduced temperature $\tau = -0.25$). The cholesteric pitch is set
to be $p = 64$, providing a slightly better resolution than the bulk
simulations.

The surface anchoring for colloids is as before: always normal, but with
$WR/K = 0.23$ and $WR/K = 2.3$. The colloid solid volume fractions are
1\% and 4\%.
For both normal and planar anchoring at the walls,
the strength is adjusted to correspond to that used for the strongly
anchoring particles.
All simulations are run for two million simulation time steps.

\noindent
\textbf{External electric field.}
For the simulations of BPIII with colloidal particles a system size of
$128^3$ is used with a cholesteric pitch of $p=32$ in lattice units. 
The simulations are
performed at chirality $\kappa=2.6$ and reduced temperature $\tau=-0.25$.  
An initial configuration of
randomly oriented and positioned double twist cylinders is used,
from which an amorphous BPIII network emerges \cite{support-oliver-bp3}. This structure
is equilibrated for $6\times10^{5}$~LB time steps until no significant
further evolution is observed. 

The colloidal particles with weak
normal anchoring ($WR / K= 0.23$), are then inserted with randomly chosen positions,
and the composite system is equilibrated for another $1.2\times10^{6}$
time steps.
At the end of the equilibration phase the uniform electric field
with reduced field strength ${\cal E} = 0.8$ 
is switched on (oriented along the $z$-direction), and the simulation
run for a further $1.2\times10^{6}$ time steps.
This is found to be sufficiently long for a hexagonal structure to
emerge and saturate.
If the electric field is then switched off, and the system quickly relaxes
to a metastable state which shows a residual anisotropy in the
colloidal distribution. The total length of this final relaxation phase
is $1\times10^6$ LB time steps.

\noindent
\textbf{Disclination rendering.}
In all cases visualisation of the defect lines is carried out via
an isosurface of the scalar order parameter~$q$ determined from the
largest eigenvalue of the local order parameter tensor. A low value
of the scalar order parameter~$q$ (typically in the range~0.12--0.14)
unambiguously identifies the disclinations.

\noindent
\textbf{Computation of the structure factor.}
To provide structural information about the field-aligned 
states and the residual anisotropy after switch off a 
procedure similar to the one described in \cite{support-oliver-bp3} is
followed.
The structure factor $S(\mathbf{k})$ in the current approach is
defined via the Fourier transform of the colloid density $F(\mathbf{k})$:
\begin{equation}
S({\bf k}) =  |F({\bf k})|^2, \mathrm{with}\
F({\bf k}) = \int d^3{\bf r} \rho({\bf r}) e^{i {\bf k. r}}.
\end{equation}
The colloid density $\rho({\bf r})$ is a discretised density and taken to be
$\rho({\bf r})=1$ if a lattice site at ${\bf r}$ is part of a colloid
and $\rho({\bf r})=0$ otherwise. This definition, which deviates from the 
standard definition of a structure factor and does not consider the colloids 
as point-like objects, allows the same structural information
to be obtained in a simple way.

\noindent
\textbf{Parameter mapping to physical units.}
To get from simulation to physical units, a calibration of scales of
length, energy and time in the simulations is required. For similar
mappings see,
e.g.~\cite{support-denniston2}.

The length scale can be set by mapping the half pitch to a realistic blue
phase unit cell size, which is in the 100--500~nm range \cite{supportblue1}. 
For bulk simulations (Fig.~1 and Fig.~3 in the main text), a suitable choice
is one where one simulation unit (lattice site) corresponds to 10~nm. Therefore
the colloidal size in Figs.~1 and~3 corresponds to $\sim$ 50~nm. 

To obtain an energy scale, it is possible to choose $A_0 \simeq 10^6$~Pa,
which is reasonable following Ref.~\cite{supportblue1} (this is also the choice in
Ref.~\cite{support-oliver2}). This choice leads to a simulation unit of
free energy (or stress) equal to $10^{8}$~Pa in Figs.~1 and~3.
From the energy and length scales one can map elastic constants to
physical units: for instance, the simulation in Fig.~1 corresponds to
a liquid crystal with splay, bend and twist (Frank) elastic constants
equal to 35 pN (or equivalently $K=70$~pN).

The timescale calibration may be obtained from (see e.g.~\cite{support-denniston})
\begin{equation}
\gamma_1=\frac{2q^2}{\mathit{\Gamma}},
\end{equation}
which relates the rotational viscosity $\gamma_1$ to the ordering strength
$q$ and the order parameter mobility $\mathit{\Gamma}$.
In the present simulations, $\mathit{\Gamma} = 0.5$ and the value of $\gamma$
for the simulation in Fig.~1 leads to $q=1/2$.   
For real liquid crystalline materials, $\gamma_1$ usually lies in range
$10^{-2}-1$ Pa s~\cite{support-deGennes}; for definiteness say
$\gamma_1 = 1$~Pa~s (equivalently 10 poise). Given the previous mapping for
free energy (or stress) units, this leads to a simulation unit of time equal
to $10^{-8}$~s. The total simulated time in Figs. 1-3 is therefore in the
millisecond range. 

To calibrate the electric field strength 
one can equate the value of the dimensionless parameter ${\cal E}$ in the 
simulations with that of a hypothetical experiment.
Given the previous value for $A_0$, and a dielectric anisotropy of 20
(or equivalently $\epsilon_\mathrm{a}=20\epsilon_0$, with $\epsilon_0$ the
dielectric permittivity of vacuum), an electric field
of 100 V~$\mu$m$^{-1}$ is obtained. This corresponds to a value of
${\cal E}\sim 0.4$ (considering a value of $\gamma=3.0857$ as in
Fig.~3 of the main text).

Given these mappings, all quantities can be easily transformed from
simulation to physical units and vice versa. For example Figs.~1 and~2
use an isotropic viscosity $\eta \simeq$~0.01 and~0.1 in simulation units,
which translate into $0.01-0.1$~Pa~s respectively. This is sensible (if 
slightly low) for a molecular nematogen in the isotropic phase. 
The effective viscosity in ordered phases is higher, but this is ensured
in our model by the coupling to the order parameter.
Note that the density is set to unity in the simulations: this
corresponds to a fluid density which is about a thousand times larger
than in experiments. This makes no difference in practice as
the Reynolds number $= \rho V\Lambda/\eta$ (where $V$ is a typical
velocity, in our simulations around $10^{-5}$ in lattice units, 
and $\Lambda$ is a typical length scale, e.g. the size of a unit cell) 
remains small enough~\cite{support-codef}. 

To summarise, the simulations represent BP-forming materials, with the
interpretation of the simulation units for length, time, and energy
density being close to 10~nm, 10~ns, and 100~MPa respectively.
For a Frank eleastic
constant of $K = 70$~pN and unit cell sizes in the range
$\lambda =$100--500~nm, a corresponding shear modulus $G \simeq
K/\lambda^2 =$ 0.3--7~kPa.

\vfill
\pagebreak

\clearpage
 
%






\end{document}